\documentclass[a4paper,11pt]{article}

\usepackage{arxiv}
\usepackage{url}
\usepackage[utf8]{inputenc} 
\usepackage{booktabs}
\usepackage{caption}
\usepackage{subcaption}
\usepackage{graphicx}
\usepackage{nameref}
\usepackage{colortbl}
\usepackage{xcolor}
\usepackage{multirow}
\usepackage{calc}
\usepackage{booktabs}
\usepackage[flushleft]{threeparttable}
\usepackage{multirow}
\usepackage{xspace}
\newcommand{\web}{Web\xspace}
\newcommand{\tor}{Tor\xspace}
\newcommand{\snpa}{SNP1\xspace}
\newcommand{\snpb}{SNP2\xspace}
\newcommand{\snpc}{SNP3\xspace}

\newcommand{\dsnpa}{DSG1\xspace}
\newcommand{\dsnpb}{DSG2\xspace}
\newcommand{\dsnpc}{DSG3\xspace}
\newcommand{\dcore}{DSGI\xspace}
\newcommand{\dunion}{DSGU\xspace}
\newcommand{\usnpa}{USG1\xspace}
\newcommand{\usnpb}{USG2\xspace}
\newcommand{\usnpc}{USG3\xspace}
\newcommand{\ucore}{USGI\xspace}
\newcommand{\uunion}{USGU\xspace}

\newcommand{\BC}{betweenness}
\newcommand{\PR}{pagerank}

\newcommand{\CC}{closeness}

\newcommand{\auth}{authscore}
\newcommand{\hub}{hubscore}
\newcommand{\eff}{efficiency}

\newcommand{\ecc}{eccentricity\xspace}
\newcommand{\tran}{transitivity\xspace}

\newcommand{\lcratio}{LCRatio\xspace}

\newcolumntype{a}{>{\columncolor{gray!10}}r}

\providecommand{\keywords}[1]
{
  \small	
  \textbf{\textit{Keywords---}} #1
}

\begin{document}

\title{\bf Onion under Microscope: An in-depth analysis of the \tor network }

\author{
        Massimo Bernaschi$^{1}$, Alessandro Celestini$^{1}$,  Marco Cianfriglia$^{1}$,\\ 
        \textbf{Stefano Guarino$^{1}$, Flavio Lombardi$^{1}$,  Enrico Mastrostefano$^{1}$} \\\\
        $^{1}$Institute for Applied Computing ``Mauro Picone'', National Research Council, Rome, Italy.\\\\
        Email: \{m.bernaschi,a.celestini,m.cianfriglia,\\s.guarino,f.lombardi,e.mastrostefano\}@iac.cnr.it
}

\date{}

\maketitle

\begin{abstract}
\tor is an anonymity network that allows offering and accessing various kinds of resources, known as hidden services, while guaranteeing sender and receiver anonymity. The \tor web is the set of web resources that exist on the \tor network, and \tor websites are part of the so-called dark web. Recent research works have evaluated \tor security, evolution over time, and thematic organization. Nevertheless, few information are available about the structure of the graph defined by the network of \tor websites. The limited number of \tor entry points that can be used to crawl the network renders the study of this graph far from being simple.
In this paper we aim at better characterizing the \tor Web by analyzing three crawling datasets collected over a five-month time frame.
On the one hand, we extensively study the global properties of the \tor Web, considering two different graph representations and verifying the impact of Tor’s renowned volatility. 
We present an in depth investigation of the key features of the \tor Web graph showing what makes it different from the surface Web graph.
On the other hand, we assess the relationship between contents and structural features.  
We analyse the local properties of the Tor Web to better characterize the role different services play in the network and to understand to which extent topological features are related to the contents of a service.
\end{abstract}


\section{Introduction}
\label{sec:introduction}

``Darkweb'' is a generic name used to denote the part of the web that is accessible only through specific privacy-preserving browsers.
Other than being non-indexed by popular search engines, the darkweb is \emph{dark} meaning that the identity of both the services offering contents and the users enjoying them may be kept anonymous by the use of overlay networks implementing suitable cryptographic protocols.
Tor is probably the best known and most widespread overlay network, owing its name to The Onion Routing protocol it is based upon.
Tor guarantees privacy and anonymity by redirecting traffic through a set of \emph{relays}, each adding a layer of encryption to the data packets they forward. Past research on the Tor network has evaluated its security~\cite{biryukov2013trawling}, evolution~\cite{Jansen:2012:MMT:2372336.2372347}, and thematic organization~\cite{spitters2014towards}.
Nevertheless, an in depth study of \tor's characteristics is difficult due to the limited number of \tor entry points on the surface \web.

In this paper, building on and extending over previous results on the topic \cite{toreros1,bernaschi2019spiders} we aim at better characterizing the \tor \web by analyzing several crawling datasets collected over a five-month time frame.
Those data are aggregated by Hidden Service (HS) (the equivalent of a domain on the surface web) in order to provide a structural analysis of these services and of their interconnections (\emph{i.e.}, hyperlinks).
In particular, the paper investigates the existence of specific \tor \web features possibly also inferring on the latent patterns of interactions among \tor users.
It also assesses whether graph metrics can be used to tell apart hidden services playing specific/unique roles in the network.
In addition, by making use of a publicly available topic-tagged dataset of Tor hidden services~\cite{al2017classifying} the paper assesses potential relationships between contents and structural features.

In line with previous work on the WWW~\cite{broder2000graph} and with a recent trend for criminal networks and dark/deep web~\cite{toreros1,onionshaveeyes,Christin:2013:TSR:2488388.2488408,griffith2017graph}, we will especially focus on the topology of the Tor web graph as a source of valuable information about Tor as a \emph{complex system}, shedding light on usage patterns as well as dynamics and vulnerabilities of the Tor network.
To that purpose, along with the three \emph{snapshot} graphs induced by the three crawling data sets, we will also consider an \emph{intersection} graph and an \emph{union} graph, in an effort to discriminate intrinsic features from noise. 
As a side effect, the present paper also addresses several open questions about the persistence of the \tor \web, showing the actual changes that took place in the quality, quantity and shape of available services and in their inter-connections over the considered time span.

This work provides several contributions, the most relevant ones summarized as follows:
\begin{itemize}

\item On the one hand, we extensively study the global properties of the \tor \web, considering two different graph representations -- directed and undirected -- and verifying the impact of \tor's renowned volatility.
Among other findings, we show that: \tor is a \emph{small world} but is \emph{inefficient}, consisting of a stable core of mostly in- and out-hubs, surrounded by an unstable periphery of services that point to or are pointed by the services in the core; when only \emph{mutual} connections are considered, we obtain an undirected version of the \tor \web that is not a small world anymore, but that better preserves the social structure of the graph, such as its community structure, which appear to be generally stable and, as such, meaningful.
Overall, \tor comes out having significant structural differences with respect to the WWW.

\item On the other hand, we analyse the local properties of the \tor \web to better characterize the role that different services play in the network and to understand to which extent topological features are related to the contents/structure of a service.
We show that authorities and hubs (as defined in~\cite{kleinberg1999authoritative}) are indeed separate services in \tor.
We provide evidence that both the volatility of Tor's hidden services and the tendency of services to cluster together are unrelated to the services' content.
We verify that switching to mutual connections also impacts on the distribution of local metrics and brings to the surface the social life of \tor in a broad sense.
Finally, we show that some topological metrics are informative of the activity occurring on a service and especially of whether this activity is ``suspicious'' (as defined in~\cite{al2019torank}).

\end{itemize}
To the best of our knowledge, this is to date the widest and most accurate study of this type on the \tor \web, exceeding previous efforts in the literature~\cite{biryukov2014content,griffith2017graph,spitters2014towards,bernaschi2019spiders}.

\subsection{Related Work}
\label{sec:related}

Interesting works studying the topology of the underlying network and/or semantically analyzing \tor contents have appeared so far.
In particular, Biryukov \emph{et al.}~\cite{biryukov2013trawling} managed to collect a large number of hidden service descriptors by exploiting a presently-fixed \tor vulnerability to find out that most popular hidden services were related to botnets.
Owen \emph{et al.}~\cite{owen2016empirical} reported over hidden services persistence, contents, and popularity, by operating 40 relays over a 6 month time frame. 
Their aim was classifying services based on their contents. 
Griffith \emph{et al.} \cite{griffith2017graph} performed a
topological analysis of the \tor hidden services graph.  They crawled the \tor
network using the \emph{scrapinghub.com} commercial service through the
\emph{tor2web} proxy onion link. 
Interestingly, they reported that more than 87\% of Darkweb sites never link to another site.  
The main difference with our work lies in both the extent of the explored network (we collected a much more extensive dataset than that
accessible through \emph{tor2web}) and the depth of the network analysis (we evaluate a far larger set of network characteristics).
Ghosh \emph{et al.}  ~\cite{Ghosh:2017:ACO:3097983.3098193} employed another automated tool to explore the \tor network and analyze the contents of onion sites for mapping onion site contents to a set of categories, and clusters \tor services to categorize onion content.  The main limitation of that work is that it focused on page contents/semantics, and did not consider network topology.  
Christin \emph{et al.}~\cite{Christin:2013:TSR:2488388.2488408} collected crawling data on Tor hidden services over an 8 month lifespan. They evaluated the
evolution/persistence of such services over time, and performed a study
on the contents and the topology of the explored network. The main difference with our work is that the \tor graph we explore is much larger, not being
limited to a single marketplace. In addition, we present here a more in depth
evaluation of the graph topology.  
De Domenico \emph{et al.}
~\cite{DeDomenico2017modeling}, used the data collected in
~\cite{Annessi2016NavigaTor} to study the topology of the \tor network.
They gave a characterization of the topology of the Darknet and proposed a
generative model for the \tor network to study its resilience.  Their viewpoint
is quite different from our own here, as they consider the network at the
autonomous system (AS) level.
Duxbury\emph{et al.}~\cite{duxbury2018network} examine the global and local network structure of an encrypted online drug distribution network. Their aim is to identify vendor characteristics that can help explain variations in the network structure. 
Their study leverages structural measures and community detection analysis to characterize the network structure. 
ToRank~\cite{al2019torank}, by Al-Nabki \emph{et al.} is an approach to rank \tor hidden services. The authors collected a large \tor dataset called DUTA-10K extending the previous Darknet Usage Text Address (DUTA) dataset~\cite{al2017classifying}. The ToRank approach selects nodes relevant to the Tor network robustness. DUTA-10K analysis reveals that only 20\% of the accessible hidden services are related to suspicious activities. It also shows how domains related to suspicious activities usually present multiple clones under different addresses.
Zabihimayvan \emph{et al.}~\cite{zabihimayvan2019broad} evaluate the contents of English Tor pages by performing a topic and network analysis on crawling-collected data composed of 7,782 pages from 1,766 unique onion domains. They classify 9 different domain types according to the information or service they host. Further, they highlight how some types of domains intentionally isolate themselves from the rest of \tor. Their measurements suggest how marketplaces of illegal drugs and services emerge as the dominant type of Tor domain. Similarly, Takaaki\emph{et al.}~\cite{takaaki2019dark} analyzed a large amount of onion domains obtained using the Ichidan search engine and the Fresh Onions site. They classified every encoutered onion domain into 6 categories, creating a directed graph and attempting to determine the relationships and characteristics of each instance.
Norbutas \emph{et al.}~\cite{norbutas2018offline} made use of publicly available crawls of a single cryptomarket (Abraxas) during 2015 and leveraged descriptive social network analysis and Exponential Random Graph Models (ERGM) to analyze the structure of the trade network.
They found out the structure of the online drug trade network to be primarily shaped by geographical boundaries, leading to strong geographic clustering, especially strong between continents and weaker for countries within Europe. As such, they suggest that cryptomarkets might be more localized and less international than thought before. 
So far, the largest \tor dataset collected from an automated \tor network exploration is due to Bernaschi \emph{et al.}~\cite{toreros1} whose work aimed at relating semantic contents similarity with \tor topology. 
Further work~\cite{bernaschi2019spiders} by the same authors features a very detailed network topology study investigating similarities and differences from surface \web and applying a novel set of measures to the data collected by automated exploration.

\subsection{Roadmap}

The rest of paper is organized as follows.
In Section~\ref{sec:dataset} we describe our dataset, including statistics about the organization of the hidden services as websites: tree map, amount of characters and links.
We introduce our hidden services graphs in Section~\ref{sec:topology}, briefly recalling how these graphs were extracted and characterizing them by analyzing properties such as global metrics, degree distribution and community structure.
In Section~\ref{sec:local} we study local (\emph{i.e.}, vertex-level) properties of our graphs in relation with content-based classification: (i) we provide a correlation analysis of several metrics, (ii) we measure the prevalence of thematic classes in graphs and communities, and (iii) we study the information gain provided by topological features for topic-based classification. 
Finally, we draw conclusions in Section~\ref{sec:conclusion}.

\section{The Dataset}\label{sec:dataset}

The dataset considered in the present paper is the result of three independent scraping procedures of the \tor \web, described in detail in~\cite{bernaschi2019spiders}. 
We executed three independent six-week runs of our customized crawler, resulting in three ``snapshots'' of the Tor \web: \snpa, \snpb and \snpc.
More details on the design of the crawler and the outcome of the scraping procedures can be found in~\cite{torMiningTool,bernaschi2019spiders}.

It is quite common to analyze a dataset obtained by crawling the web.
Yet, it must be kept in mind that the analysis may be susceptible to fluctuations due to the order in which pages have been first visited -- and, hence, not revisited thereafter~\cite{Lehmberg:2014:GSW:2615569.2615674}.
In the case of the \tor \web, the issue is exacerbated by the renowned volatility of \tor hidden services~\cite{biryukov2013trawling,biryukov2014content,owen2016empirical}.
By executing three independent scraping attempts over five months, we aimed at making our analysis more robust and at telling apart ``stable'' and ``temporary'' features of the \tor \web.

\begin{figure}[ht]
   \centering
	\includegraphics[width=0.65\textwidth]{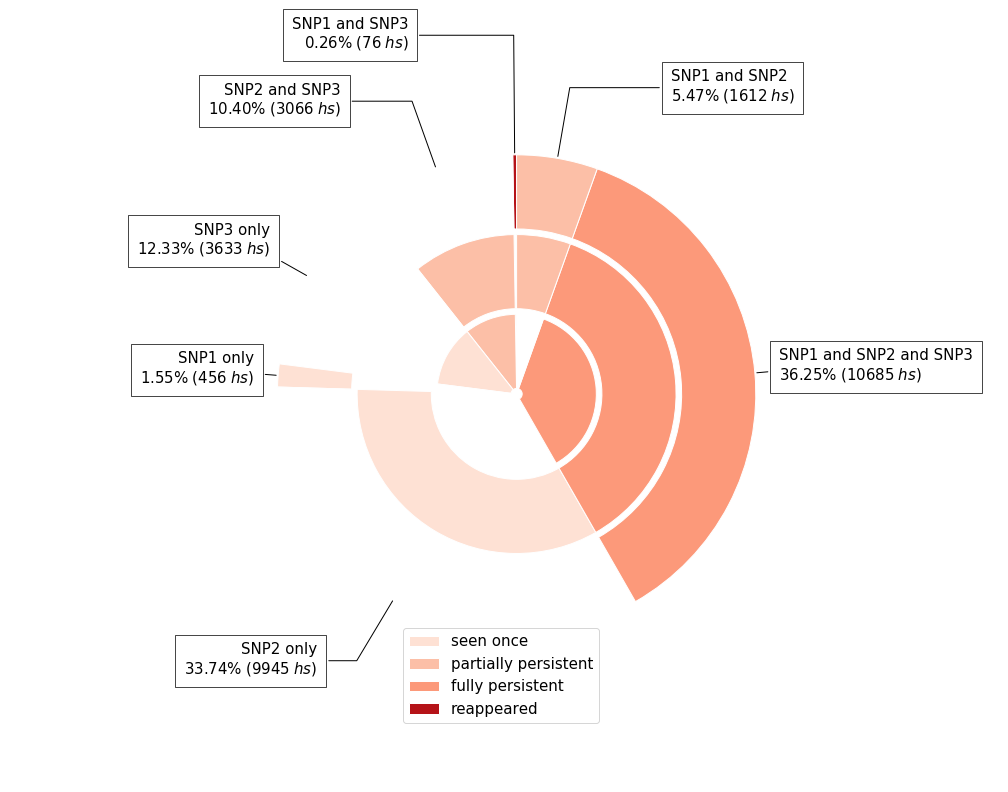}
	\caption{Services persistence over time: outer disc is \snpa, middle disc is \snpb, inner disc is \snpc.}
	\label{fig:services_persistance}
\end{figure}

In total, we reached millions of onion pages (more than 3 millions in the second run alone) and almost 30 thousands different hidden services.
The distribution of these hidden services across the three snapshots is depicted through a pie-diagram in Figure \ref{fig:services_persistance}.
Albeit active services may temporarily appear offline to the crawler (\emph{e.g.}, due to all paths to those services being concurrently unavailable), these statistics are quite informative about the volatility of the Tor web.
``Only'' 10685 onion URLs were successfully reached by all three crawling runs. It is quite likely that those  hidden services were durably present over the considered five months time frame; they account for, respectively, 83.3\% of \snpa, 42.2\% of \snpb and 61.2\% of \snpc.
Among the hidden services that are absent in just one of the three data sets, especially notable are the 76 hidden services that reappeared in \snpc after not having been found during \snpb.
To gain a better understanding of the structure and degree of mutability of Tor contents, we resorted to a few high-level indicators.
On the one hand, we reconstructed the whole tree-structure of each and every hidden service sub-domain and pages; on the other hand, we computed the total number of characters and the total number of hyperlinks (\emph{i.e.}, number of \emph{href}s in the HTML source) of each service.
These metrics can be used as proxies for the complexity of Tor domains, taking both structure and contents into account.
In particular, we identified all onions whose tree structure and/or text volume remained constant across all snapshots both to obtain a further measure of volatility and to better characterize stable services. 

\begin{figure*}[htbp]
 \centering
        \includegraphics[width=.6\textwidth]{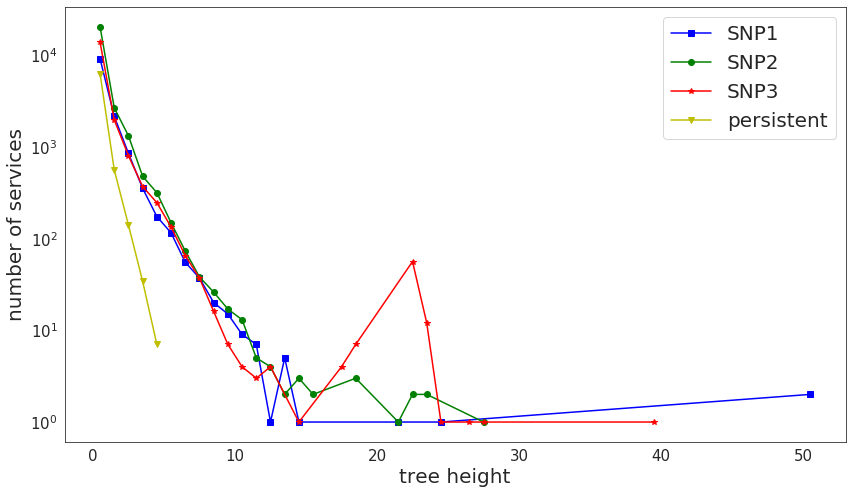}
    \caption{The statistical distribution of tree heights.}
    \label{fig:tree}
\end{figure*}

\begin{figure}[htbp]
 \centering
        \includegraphics[width=.54\textwidth]{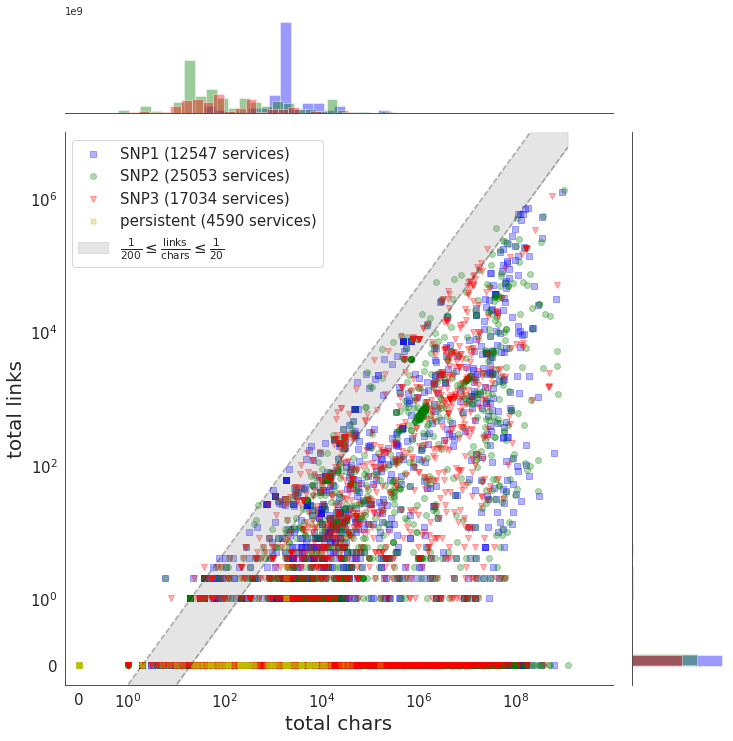}
        \caption{The joint distribution of char and link counts. Marginal distributions are reported externally to the main box.}
    \label{fig:chars_links}
\end{figure}

Figure~\ref{fig:tree} shows the statistical distribution of tree heights for the three snapshots and for the 6961 hidden services -- approximately 65\% of all stable onion URLs -- whose tree structure remains consistent across all snapshots (denoted ``persistent'' in the legend).
The figure shows that the trees are generally very short -- thus poorly informative on the ``nature'' of the service -- and that only ``short'' trees seem to be persistent.
Figure~\ref{fig:chars_links} instead shows the joint distribution of char and link counts, where, together with the three snapshots, we considered the 4590 hidden services  -- approximately 43\% of all stable onion URLs -- whose char count remains constant across all snapshots (again, denoted ``persistent'' in the legend).
While the char count is generally variable, services with 0 links are predominant.
This is especially true in hidden services with a persistent char count -- hence, presumably persistent contents -- that appear to be very peripheral, other than having a char count lower than average.  
Although there is no visible correlation between char and link counts, there seems to be an almost constant upper bound for the ratio of links over chars.
This is not entirely surprising, since clickable hrefs require a small amount of text.
In particular, we highlighted the plane region bounded by web pages having one link every 20 chars (\emph{i.e.}, $\approx3$ words) and one link every 200 chars (\emph{i.e.}, 1 to 2 sentences), a pattern that we may expect for link directories. 
In Section~\ref{sec:local}, we will correlate the \emph{links-to-char ratio} (LCRatio) with a set of topological features for better investigating the relation of topological and content-related feature.
The above-reported analysis also leads to a consequent finding: automatically detecting hidden services that stay durably online but with different names is hardly feasible, because often the contents vary with time even for stable hidden services.
The tree structure and the char count are in fact only proxies for the contents of a hidden service, and yet the portion of stably-structured as well as that of stably-sized hidden services is far from 100\%.
Albeit we ideally aimed at determining the persistence of a service regardless of whether it changed its name -- \emph{e.g.}, to prevent from being tracked down -- in the remainder we will therefore rely on the onion URL as the unique hidden service identifier.


\section{Characterizing the Tor Web Graphs}\label{sec:topology}

The aforementioned scraping procedures produced three WARC files.
From each of them, we extracted two graphs: a \emph{Directed} Service Graph (DSG) and an \emph{Undirected} Service Graph (USG).
As detailed in~\cite{torMiningTool}, a vertex of these graphs represents the set of pages belonging to a hidden service.
In the DSG a directed edge is drawn from hidden service HS1 to HS2 if any page in HS1 contains, at least, a hypertextual link to any page in HS2\footnote{Edges from/to the surface web have been ignored.}.
The directed graphs obtained from the three snapshots are denoted \dsnpa, \dsnpb, and \dsnpc, respectively.
In the USG, instead, an undirected edge connects hidden services HS1 and HS2 if they are \emph{mutually} connected in the corresponding DSG, that is, if there exists at least one page in HS1 linking any page in HS2 \emph{and} at least one page in HS2 linking any page in HS1.
A vast majority of isolated vertices ``naturally'' emerges from only considering mutual connections; these are ignored in the following -- \emph{i.e.}, we consider edge-induced graphs -- since they convey no structural information.
The undirected graphs obtained from the three snapshots are denoted \usnpa, \usnpb, and \usnpc, respectively.

As highlighted in Section~\ref{sec:dataset}, the snapshot graphs are inevitably conditioned by the effect of scraping a reputedly volatile network.
We therefore also consider the intersection and the union of the aforementioned graphs.
Precisely, we consider edge-induced intersection and union, because hyperlinks are generally ``less stable'' than hidden services (for more details, see~\cite{bernaschi2019spiders}) and, again, we choose to ignore isolated vertices. 
This means that \dcore is induced by the set of common edges among \dsnpa, \dsnpb and \dsnpc, whereas \dunion is induced by the set of edges existing in, \emph{at least}, one of them.
Analogously, \ucore is induced by the set of common edges among \usnpa, \usnpb and \usnpc, whereas \uunion is induced by the set of edges existing in, \emph{at least}, one of them.
Both \ucore and \uunion are composed by multiple connected components.

We do not preserve multi-edges so as to permit a direct comparison with most previous work on other web and social/complex networks -- indeed, even simple metrics (\emph{e.g.}, the degree) may be misleading for multi-graphs.
However, algorithms on web graphs are often based -- either directly or indirectly -- on modeling a \emph{random web surfer}~\cite{page1999pagerank}, so we shall not neglect the fact that some edges may be more likely travelled by the surfer than others.
In both directed and undirected graphs, we store the information about the number of links that have been ``flattened'' onto an edge as a \emph{weight} attribute assigned to that edge -- taking the minimum available weight for edges of our intersection graph and the maximum for the union.
Whether to use this weight in practice will be separately discussed for each task.
We believe that the edge weight may be interpreted as a measure of connection strength, expressing endorsement/trust but \emph{not} altering distances.
This is the reason why we will ignore it in all global metrics discussed in the following.

\subsection{Global Metrics}\label{sec:global}

As a first step towards a possible understanding of \tor dynamics, we characterize our ten graphs relying on well-known metrics, summarized in Table~\ref{tab:global_metrics} -- while graph-related symbols used throughout the paper are reported in Table~\ref{tab:notation}. 
Many of such metrics are undefined for disconnected graphs, or may provide misleading results when evaluated over multiple isolated components.
To make up for it and allow for a fair comparison, in the remainder of this section we only consider the giant (weakly) connected component of all disconnected graphs.
It is worth mentioning that the three DSGs, and therefore their union \dunion, consist of a single weakly connected component.
On the contrary, all USGs are weakly disconnected graphs.
\dcore is also disconnected, albeit only two hidden services -- violet77pvqdmsiy.onion and typefacew3ijwkgg.onion -- are isolated from the rest and connected by an edge.
We will again consider the graphs in their entirety in Section~\ref{sec:clustering}. 
\begin{table}[htbp]
\caption{Basic graph notations and definitions used throughout the paper.}
\label{tab:notation}
\centering
\begin{tabular}{llll}
\toprule
\textbf{Symbol} & \textbf{Definition} \\
\midrule
  \multirow{2}{*} {$G = (V, E)$} & Graph with vertex set $V$& \multirow{2}{*} {$\sigma_{vu}(t)$}&Number of shortest paths from $v$ to $u$\\
	& and edge set $E$& & including $t$\\
$N$             & Number of nodes: $N=|V|$ & $\sigma_{vu}$ & Number of shortest paths from $v$ to $u$ \\
$M$              & Number of edges: $M=|E|$ & $\mathrm{dist}(v,u)$  & Shortest path length from $v$ to $u$\\
\bottomrule
\end{tabular}
\end{table}
\begin{table}[htbp]
\caption{Global metrics notations and definitions.}
\label{tab:global_metrics}
\centering
\begin{tabular}{ll}
\toprule
\textbf{Symbol} & \textbf{Definition} \\
\midrule
\multicolumn{2}{c}{Global metrics valid for both DSG and USG} \\
\midrule
$\langle\deg\rangle$ & Average (in-/out-) degree \\
$\rho$               & Assortativity: see (26) in \cite{newman2003mixing}\footnotemark \\
$d$                  & Diameter: $\max_{v\in V} \max_{u\in V} \mathrm{dist}(v,u)$ \\
$\langle \mathrm{dist}\rangle$ & Average shortest path length \\
$E_{glo}$  & Global efficiency: $\frac{1}{N(N-1)} \sum_{u\neq v\in V} \frac{1}{d(u,v)}$ \\
\midrule
\multicolumn{2}{c}{Global metrics valid for DSG only} \\
\midrule
$\Delta_{in}/N$  & Normalized maximum in-degree  \\
$\Delta_{out}/N$ & Normalized maximum out-degree \\
$\mathrm{Cen}_{out}$ & Out-degree centralization: $\frac{ N*\Delta_{out} - \sum_{v\in V} \deg_{out}(v)}{(N-1)^2}$ \\
$T$   & Global transitivity: $\frac{\# (u,v,w):\; u \to v \wedge u \to w \wedge (v \to w \vee w \to v)}{\# (u,v,w):\; u\to v \wedge u \to w}$ \\ 
\midrule
\multicolumn{2}{c}{Global metrics valid for USG only} \\
\midrule
$\Delta/N$ & Normalized maximum degree \\
$\mathrm{Cen}$ & Degree centralization: $\frac{N*\Delta - \sum_{v\in V} \deg(v)}{(N-1)(N-2)}$ \\
$C$          & Global clustering coefficient: $\frac {\mbox{\# closed triplets}}{\mbox{\# all triplets}}$ \\
\bottomrule
\end{tabular}
\end{table}
\footnotetext{Beware that $\rho$ is defined differently for directed and undirected graphs.}

Table~\ref{tab:global_directed} provides a first glimpse of the structure of the \tor web in terms of global metrics of our DSGs. 
We observe a significant variance in the sizes $N$ and $M$ of the three snapshots, which is however consistent with publicly available aggregated statistics\footnote{https://metrics.torproject.org/hidserv-dir-onions-seen.html?start=2017-01-01\&end=2017-05-01}, as already discussed in~\cite{bernaschi2019spiders}.
In all graphs, there are huge out-hubs reaching 35\% to 61\% of the network, but no equivalently prominent in-hubs.
The values of $\mathrm{Cen}_{\mathrm{out}}$ show that the network tends to be out centralized and the high values of $\Delta_{out}/N$ denotes the importance of large out-hubs in the graphs' connectivity.
However, the greatest of such hubs emerges in the largest graphs, and $\Delta_{in}/N$ and $\Delta_{out}/N$ are comparably smaller in \dcore with respect to the snapshots, 
suggesting that the degree of such stable hubs is heavily influenced by the non-persistent nodes. Our analysis suggests the presence of a stable core mainly composed by in- and out-hubs, and of an unstable periphery composed of nodes that point to or are pointed by the nodes in the core.   
The assortativity $\rho$ is computed according to Newman's original definition~\cite{newman2003mixing}, \emph{i.e.}, it measures the correlation between a  node's out-degree and the adjacent nodes' respective in-degree.
All networks are \emph{disassortative}, meaning that links are more likely to connect high-out-degree nodes to low-in-degree nodes, or low-out-degree nodes to high-in-degree nodes.
In other words, hubs (\emph{i.e.}, link directories) ``rarely'' link to authorities. 
The reported transitivity $T$ measures how often vertices that are adjacent to the same vertex are connected in, at least, one direction.
This happens more often than expected, as it emerges comparing $\frac{\langle \deg \rangle}{N}$ with the measured values for $T$.
The diameter $d$ and the average path length $\langle \mathrm{dist} \rangle$ are approximately logarithmic in $N$, as in most social and web graphs.
In other words, the Tor web graph is a \emph{small world}.
However, these metrics only consider \emph{finite} (\emph{i.e.}, existing) paths.
If we also include \emph{infinite} paths, as done when computing the global efficiency $E_{glo}$, we see that many vertex pairs are disconnected and information diffusion in the Tor web is quite inefficient.

\begin{table}[htbp]
\centering
\caption{Global metrics for the directed service graphs.}
\label{tab:global_directed}
\begin{tabular}{lrrrrr}
\toprule
{} &    \dsnpa &     \dsnpb &     \dsnpc &    \dcore &    \dunion \\
\midrule
$N$                               & 12829     &  25308     &  17460     &  7669     &  29473     \\
$M$                               & 72556     & 113014     & 103402     & 28913     & 187415     \\
$\langle \mathrm{dist} \rangle$   &     3.793 &      4.96  &      3.665 &     3.983 &      3.821 \\
$d$                               &    10     &     12     &     10     &    10     &      9     \\
$E_{\mathrm{glo}}$                &     0.011 &      0.008 &      0.041 &     0.007 &      0.029 \\
$\langle \deg \rangle$            &     5.656 &      4.466 &      5.922 &     3.77  &      6.359 \\
$\frac{\Delta_{\mathrm{in}}}{N}$  &     0.016 &      0.01  &      0.084 &     0.007 &      0.05  \\
$\frac{\Delta_{\mathrm{out}}}{N}$ &     0.437 &      0.508 &      0.611 &     0.348 &      0.566 \\
$\rho$                            &    -0.319 &     -0.327 &     -0.162 &    -0.374 &     -0.168 \\
$\mathrm{Cen}_{\mathrm{out}}$     &     0.436 &      0.508 &      0.61  &     0.348 &      0.566 \\
$T$                               &     0.004 &      0.002 &      0.002 &     0.004 &      0.002 \\
\bottomrule
\end{tabular}
\end{table}

Now, we switch to the giant connected components of the USGs.
From Table~\ref{tab:global_undirected} we immediately see that the sizes $N$ and $M$ of the three snapshots are again variable, but, somewhat surprisingly, \usnpc is now the biggest.
This is probably related to the existence in \usnpc of a large hub that is absent in the other two snapshots.
Indeed the value of $\mathrm{Cen}$ shows that the network tends to be centralized and the high value of $\Delta/N$ denotes the importance of such large out-hubs in the graphs' connectivity.
Albeit all of these graphs are much smaller, the average distance and the diameter are comparable to, or even greater than, their directed counterparts.
This tells us that long ``bidirectional'' paths exist and that, when only mutual connections are considered, the Tor web is not a \emph{small world} anymore.
For undirected graphs, the assortativity $\rho$ measures the tendency of a node to connect with other nodes having similar degree.
As all networks are again \emph{disassortative}, we know that hubs are more likely connected with peripheral nodes.
However, in both \usnpa and \usnpb, where the tendency of vertices to cluster together is not hidden by the presence of a dominant hub, we notice that the clustering coefficient $C$ is one order of magnitude greater than $\frac{\langle \deg \rangle}{N}$, that is, its estimate in completely random graphs.
Finally, these graphs are much more efficient than the DSGs, albeit this may be due simply to all pairs now being connected in both directions (in fact, $E_{glo}\approx \frac{1}{\langle \mathrm{dist} \rangle}$).

\begin{table}[htbp]
\centering
\caption{Global metrics for the undirected service graphs.}
\label{tab:global_undirected}
\begin{tabular}{lrrrrr}
\toprule
{} &  \usnpa &  \usnpb &   \usnpc &  \ucore &  \uunion \\
\midrule
$N$                             & 208     & 225     & 2084     &  87     & 2244     \\
$M$                             & 398     & 467     & 2289     & 143     & 2685     \\
$\langle \mathrm{dist} \rangle$ &   4.301 &   3.941 &    2.707 &   3.939 &    2.881 \\
$d$                             &  15     &   9     &   10     &   9     &   11     \\
$E_{\mathrm{glo}}$              &   0.285 &   0.295 &    0.408 &   0.308 &    0.389 \\
$\langle \deg \rangle$          &   3.827 &   4.151 &    2.197 &   3.287 &    2.393 \\
$\frac{\Delta}{N}$              &   0.188 &   0.213 &    0.7   &   0.23  &    0.65  \\
$\rho$                          &  -0.077 &  -0.121 &   -0.602 &  -0.023 &   -0.489 \\
$\mathrm{Cen}$                  &   0.171 &   0.197 &    0.699 &   0.197 &    0.649 \\
$C$                             &   0.203 &   0.208 &    0.001 &   0.259 &    0.002 \\
\bottomrule
\end{tabular}
\end{table}

\subsection{Degree Distribution}\label{sec:degree_distribution}

\begin{figure*}[htbp]
    \centering
    \begin{subfigure}[b]{0.4\textwidth}
        \includegraphics[width=\textwidth]{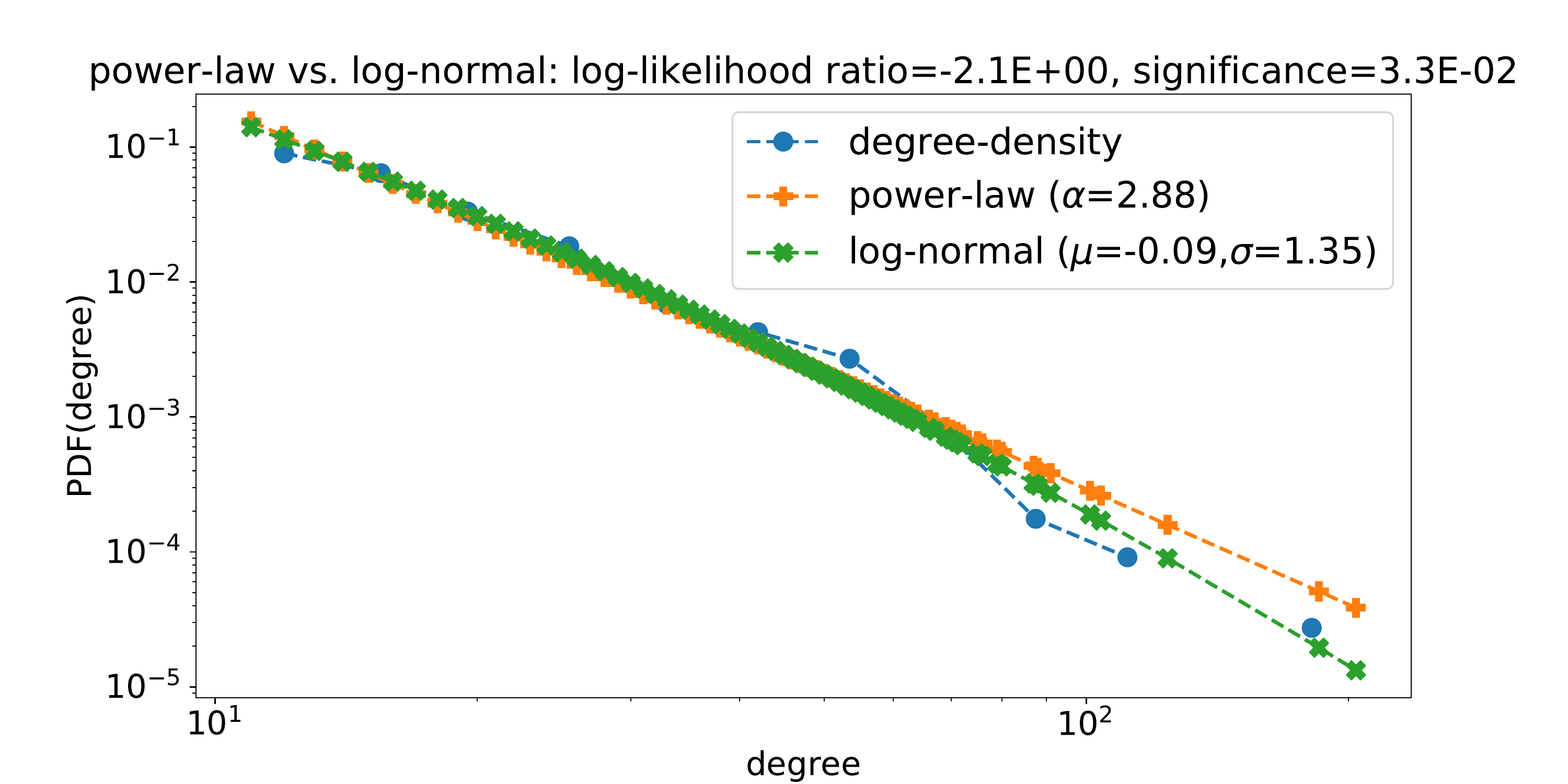}
        \caption{\dsnpa in-degree.}
	    \label{fig:dsg1_indegree}
    \end{subfigure}
    \begin{subfigure}[b]{0.4\textwidth}
        \includegraphics[width=\textwidth]{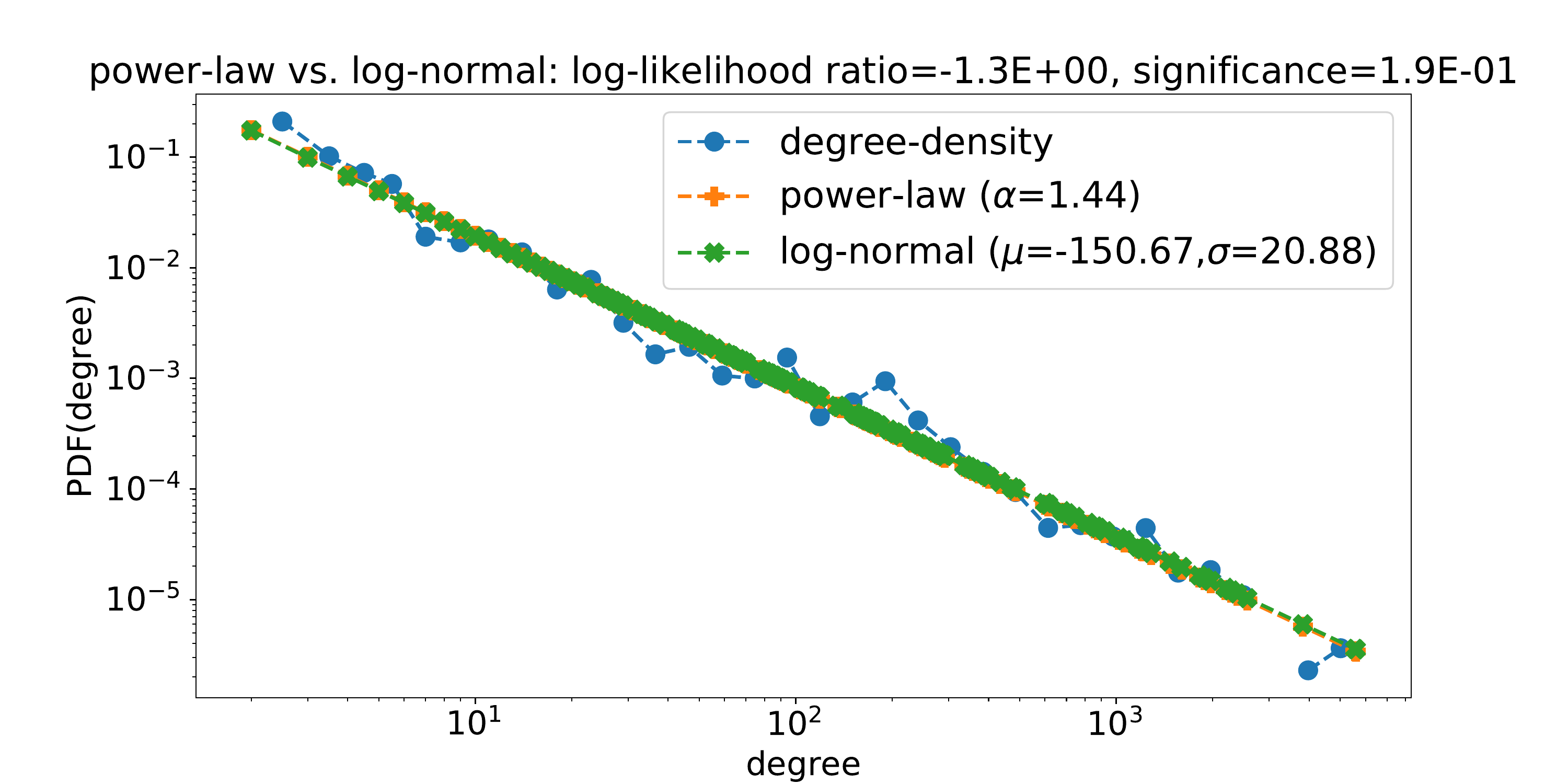}
	    \caption{\dsnpa out-degree.}
	    \label{fig:dsg1_outdegree}
    \end{subfigure}
    
    \begin{subfigure}[b]{0.4\textwidth}
        \includegraphics[width=\textwidth]{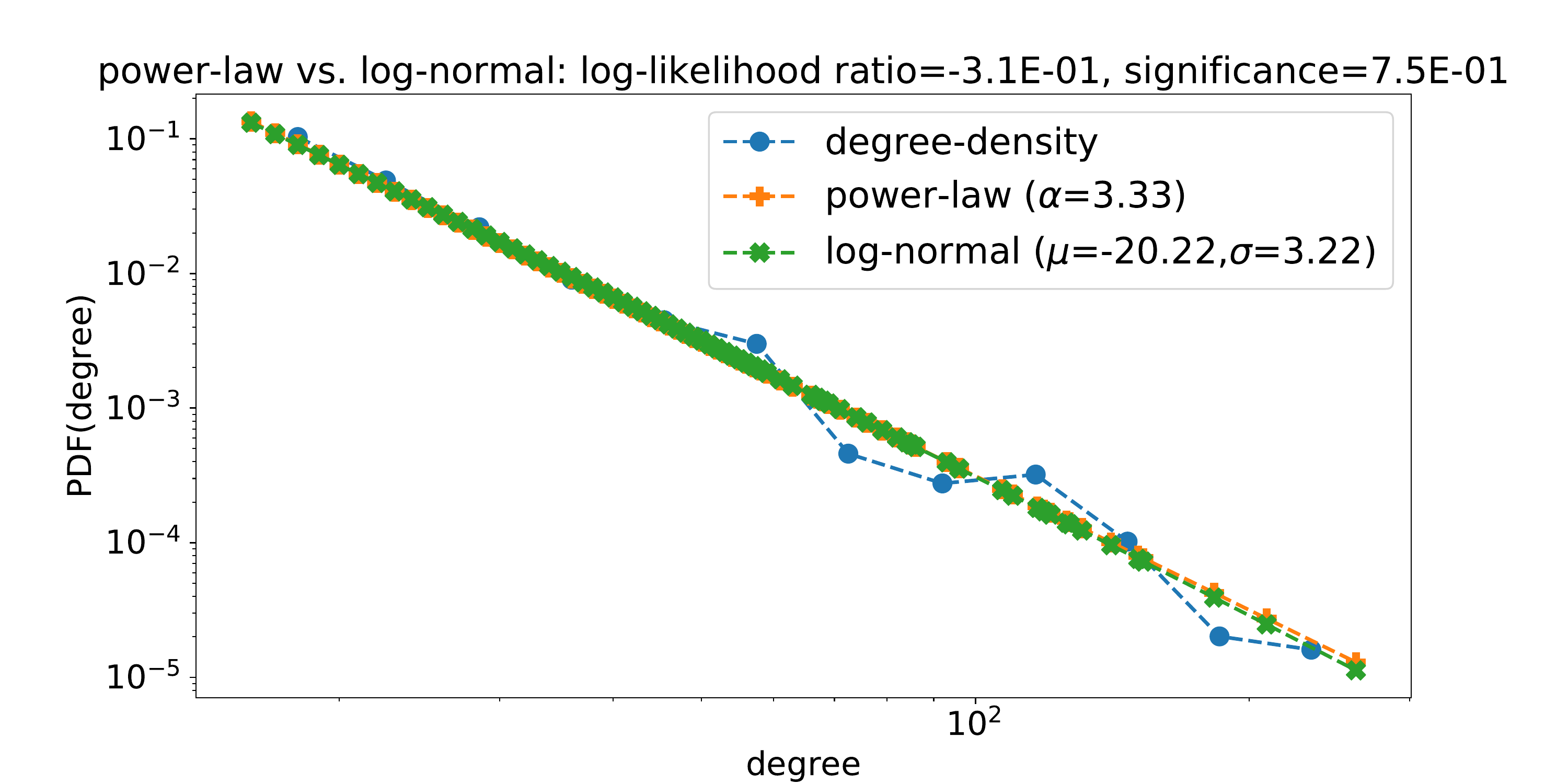}
        \caption{\dsnpb in-degree.}
	    \label{fig:dsg2_indegree}
    \end{subfigure}
    \begin{subfigure}[b]{0.4\textwidth}
        \includegraphics[width=\textwidth]{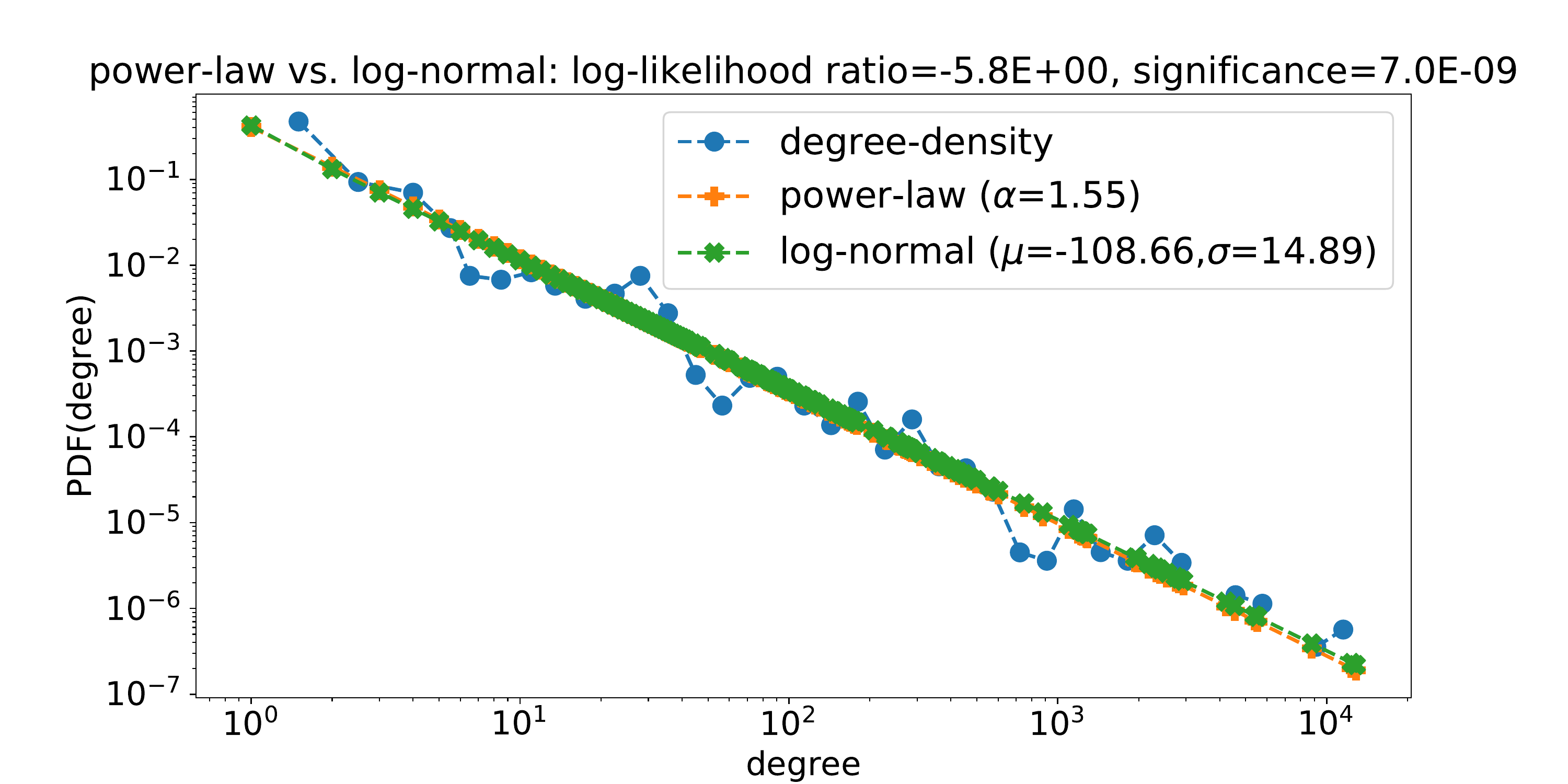}
	    \caption{\dsnpb out-degree.}
	    \label{fig:dsg2_outdegree}
    \end{subfigure}
    
    \begin{subfigure}[b]{0.4\textwidth}
        \includegraphics[width=\textwidth]{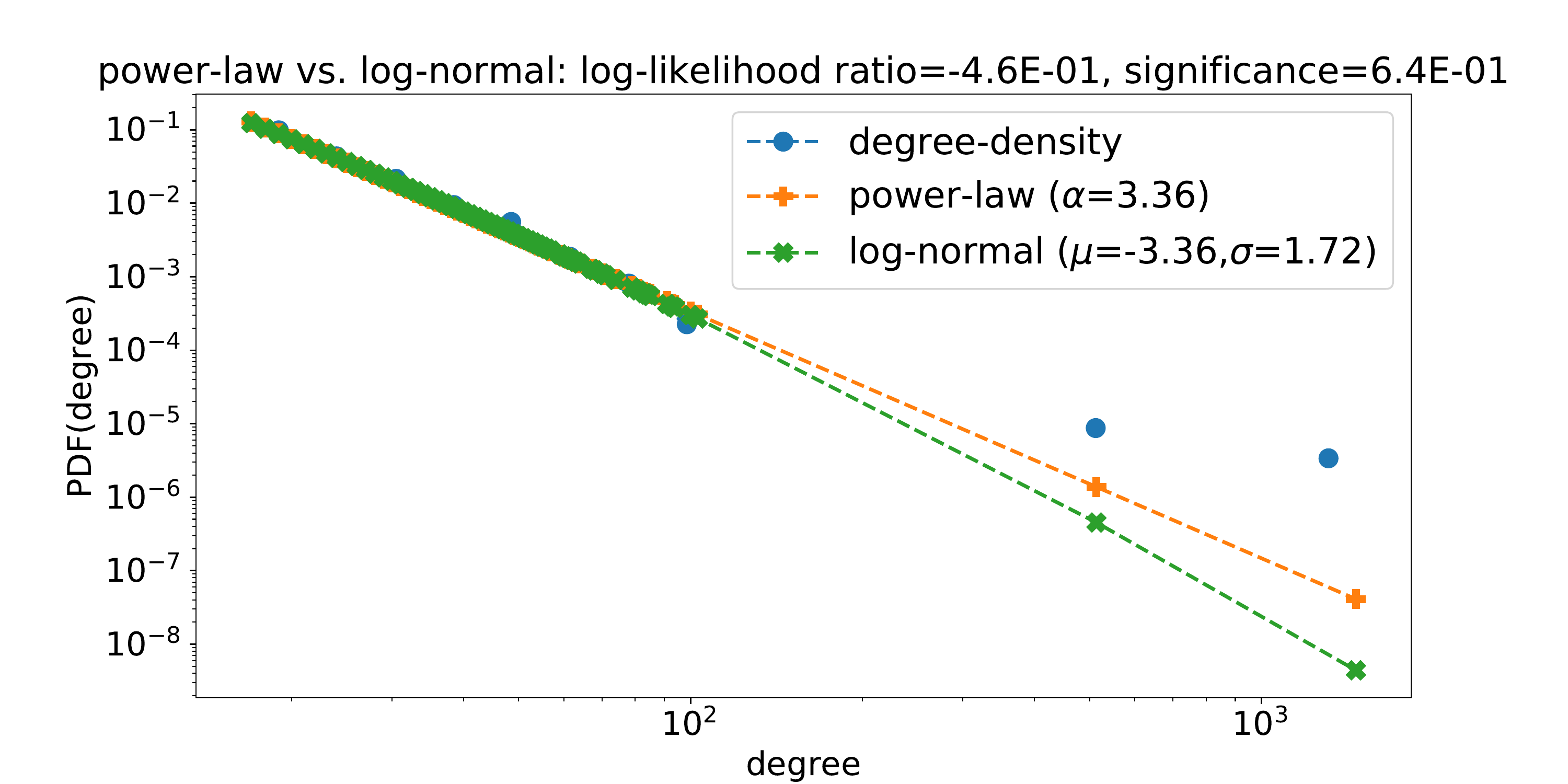}
        \caption{\dsnpc in-degree.}
	    \label{fig:dsg3_indegree}
    \end{subfigure}
    \begin{subfigure}[b]{0.4\textwidth}
        \includegraphics[width=\textwidth]{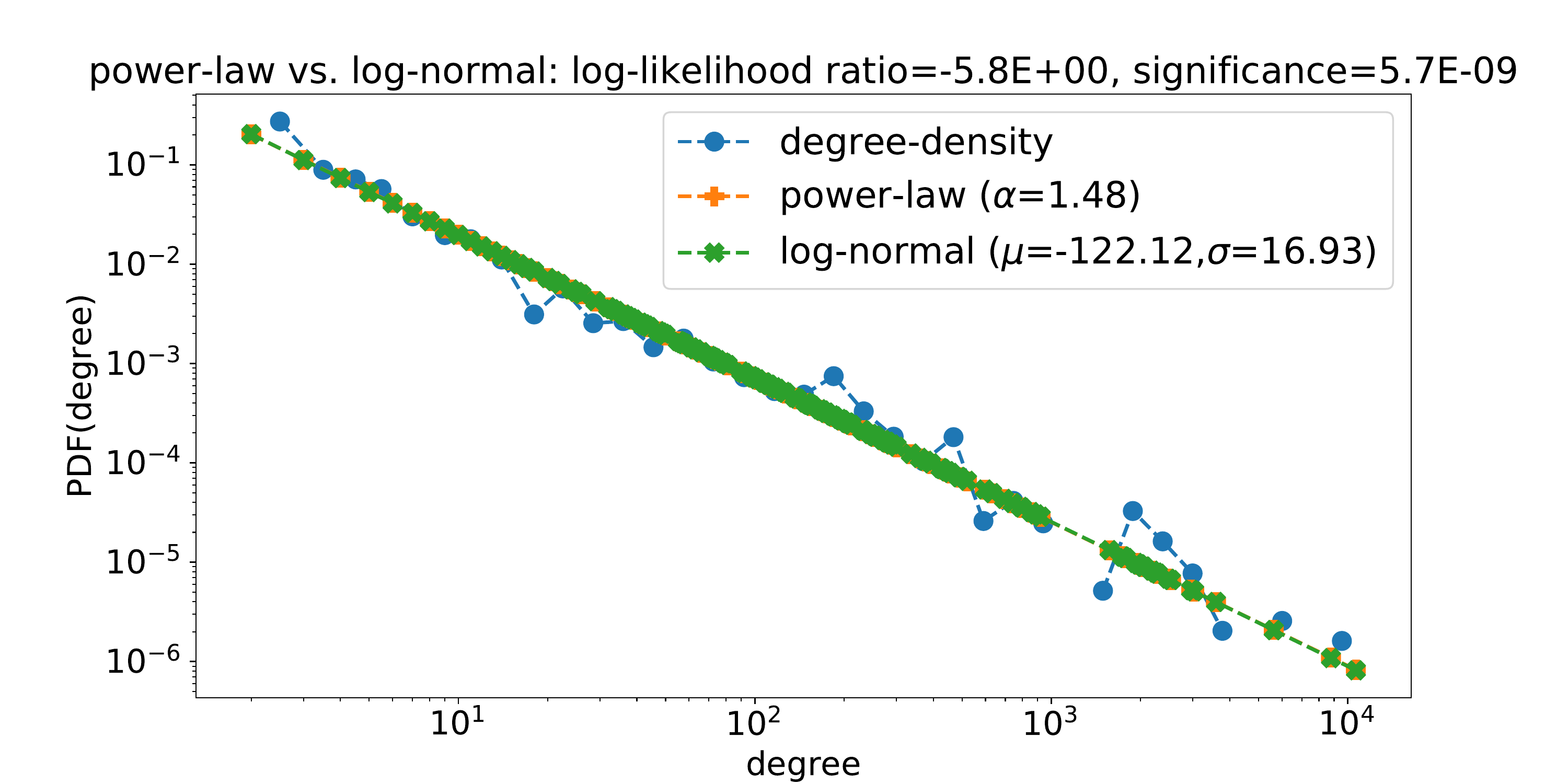}
	    \caption{\dsnpc out-degree.}
	    \label{fig:dsg3_outdegree}
    \end{subfigure}
    
    \begin{subfigure}[b]{0.4\textwidth}
        \includegraphics[width=\textwidth]{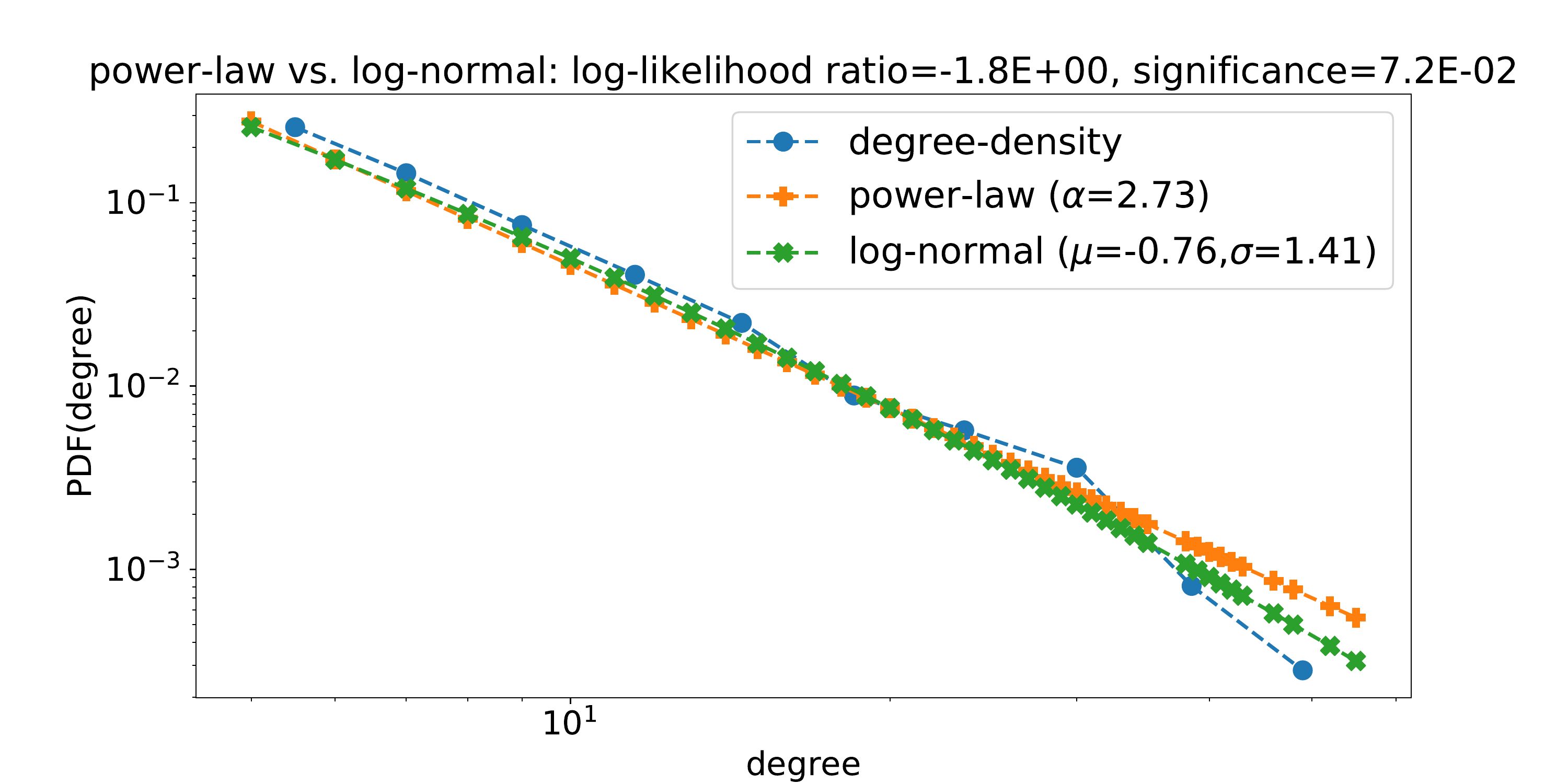}
        \caption{\dcore in-degree.}
	    \label{fig:dsgi_indegree}
    \end{subfigure}
    \begin{subfigure}[b]{0.4\textwidth}
        \includegraphics[width=\textwidth]{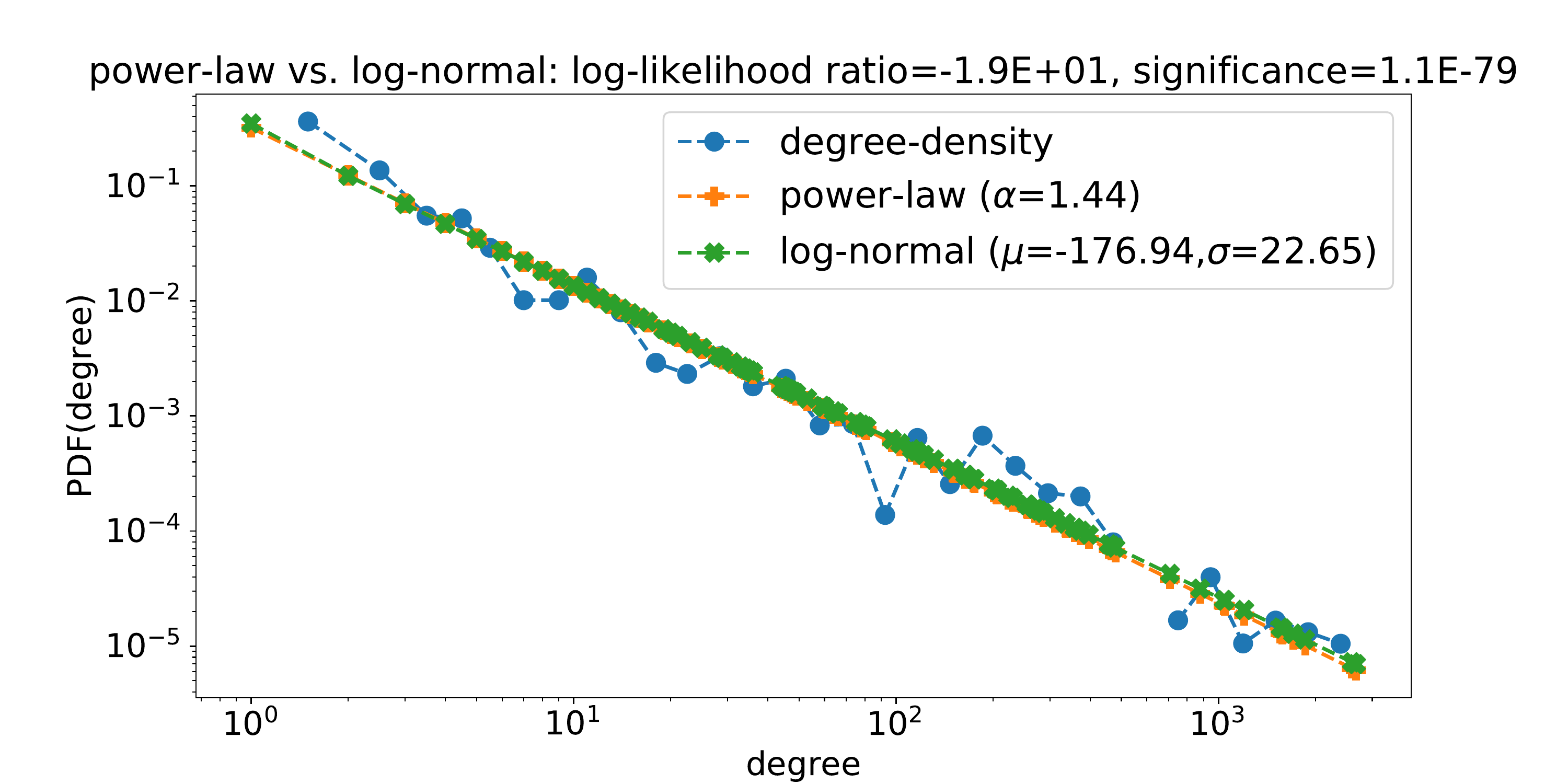}
	    \caption{\dcore out-degree.}
	    \label{fig:dsgi_outdegree}
    \end{subfigure}
    
    \begin{subfigure}[b]{0.4\textwidth}
        \includegraphics[width=\textwidth]{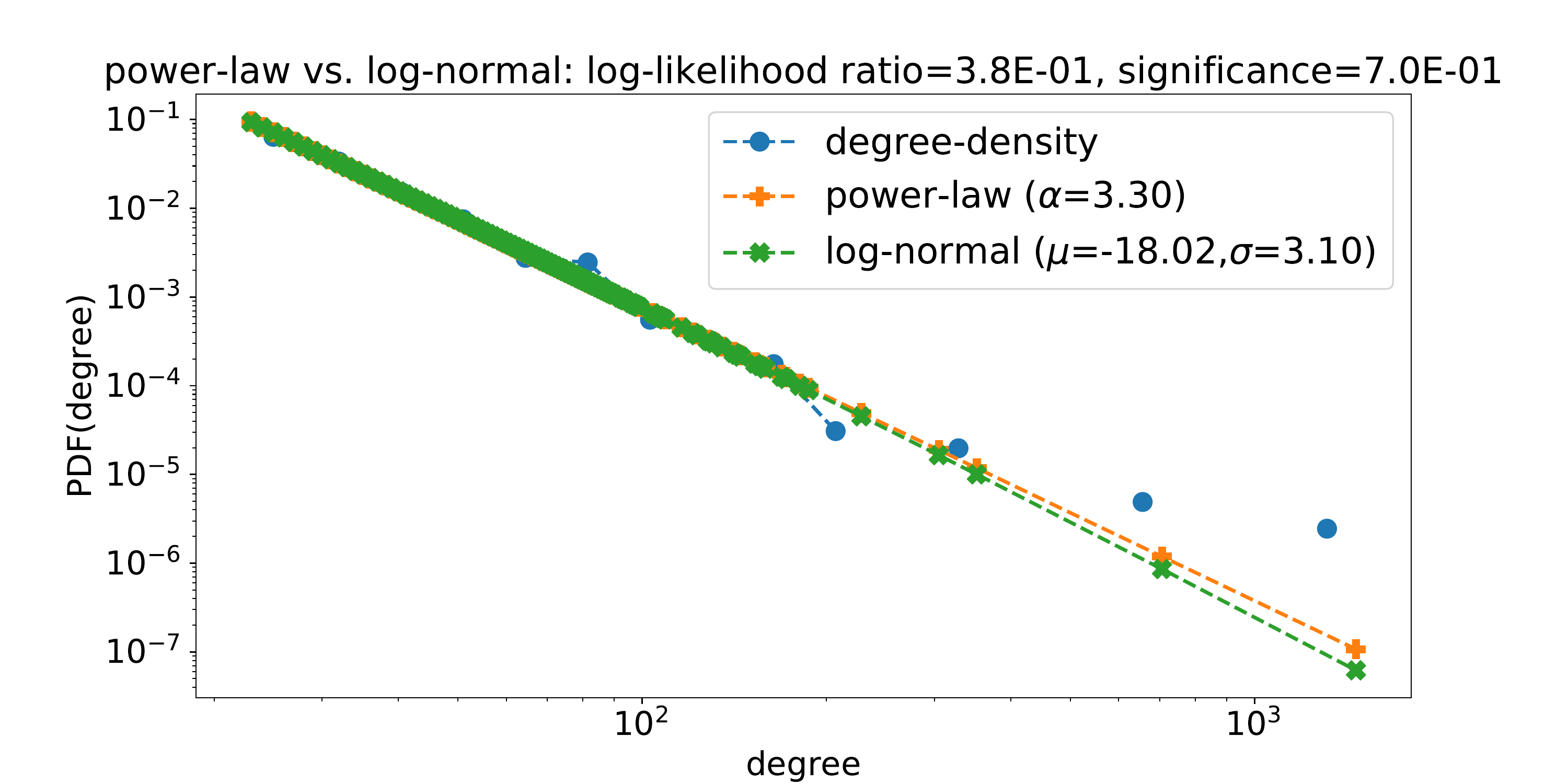}
        \caption{\dunion in-degree.}
	    \label{fig:dsgu_indegree}
    \end{subfigure}
    \begin{subfigure}[b]{0.4\textwidth}
        \includegraphics[width=\textwidth]{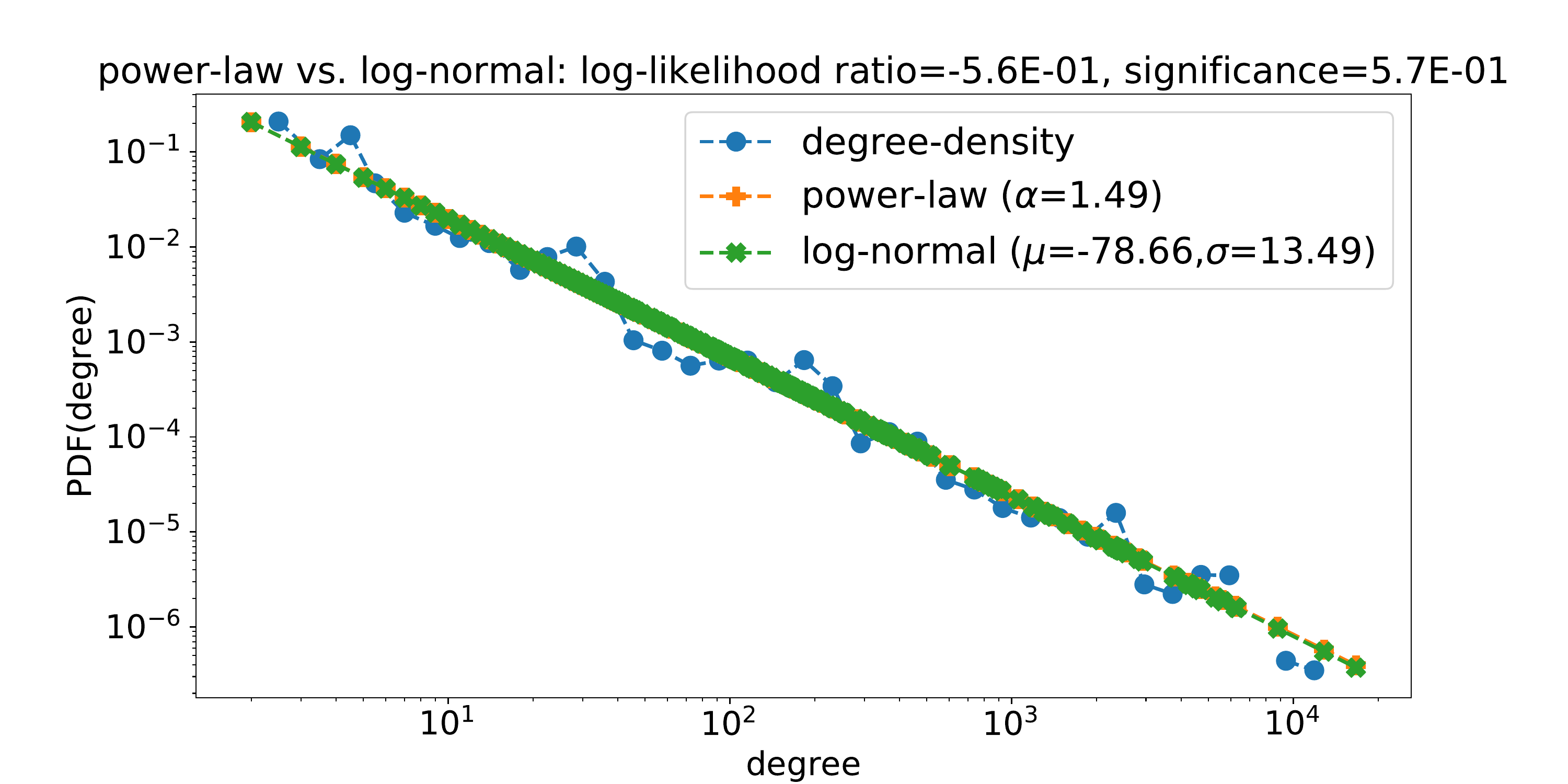}
	    \caption{\dunion out-degree.}
	    \label{fig:dsgu_outdegree}
    \end{subfigure}
    \caption{The degree distribution for the DSGs.}
    \label{fig:degree_directed}
\end{figure*}

Figure~\ref{fig:degree_directed} shows the distributions of the in- and out-degree for all five directed graphs on a log-log scale.
We fitted a power-law to the distribution using the statistical methods developed in~\cite{clauset2009power}, relying on the implementation provided by the \textsc{powerlaw} python package~\cite{alstott2014powerlaw}.
We also fitted a log-normal and reported the comparison of these two fits in the figure.
As already suggested in the literature~\cite{mitzenmacher2004brief}, a log-normal distribution may be a better fit of degree distributions in many complex networks.
In particular, a recent work suggests that a log-normal distribution may emerge from the combination of preferential attachment and growth~\cite{sheridan2018preferential}.
In our DSGs, the log-normal fit slightly outperforms the power-law fit for the tail of the distribution.
It is worth specifying that \textsc{powerlaw} autonomously finds a lower-bound $k_{\min}$ for degrees to be fitted.
In this case, even if $k_{\min}$ is much less than the maximum degree, all values greater than $k_{\min}$ account for just a small percentage of the whole graph.
However, we believe this should not prevent from taking these fits seriously into consideration: the tail of the distribution \emph{de facto} describes the central part of the graph that actually has a meaningful structure -- as opposed to the bulk of the distribution mostly depicting vertices with out-degree 0 (83\% to 95\% of the graph according to the specific DSG considered) and/or in-degree 1 (17\% to 43\%).

Looking at the numbers, we immediately realize the huge difference between the in-degree and out-degree distribution.
This is not surprising, since very different interpretations can be given of these two quantities: while the in-degree measures the ability of a service to attract interest, the out-degree is just a measure of a ``design choice'' -- how many links to include in a web page -- which is related to the nature of the service but not inherently associated to any social characteristics of the network. 
For \dsnpb, \dsnpc and \dunion the $\alpha$ exponent of the in-degre distribution is greater than the threshold 3 that is known to control the variance of the distribution, whereas for \dsnpa $\alpha$ is close to 2.9 and for the \dcore graph it is close to 2.7.
Intuitively, this says that the \dsnpa and the \dcore ``look more like'' social networks.
All out-degree distributions have instead $\alpha\approx1.5$, which is a very low value reflecting the existence of many large out-hubs -- \emph{i.e.}, link directories or similar web services.
In other words, the long tail of the out-degree distribution says that the large value of $\Delta_{out}$ observed in Section~\ref{sec:global} is not an isolated case, but rather an evidence of a general trend.

\begin{figure*}[ht]
    \centering
    \begin{subfigure}[b]{0.4\textwidth}
        \includegraphics[width=\textwidth]{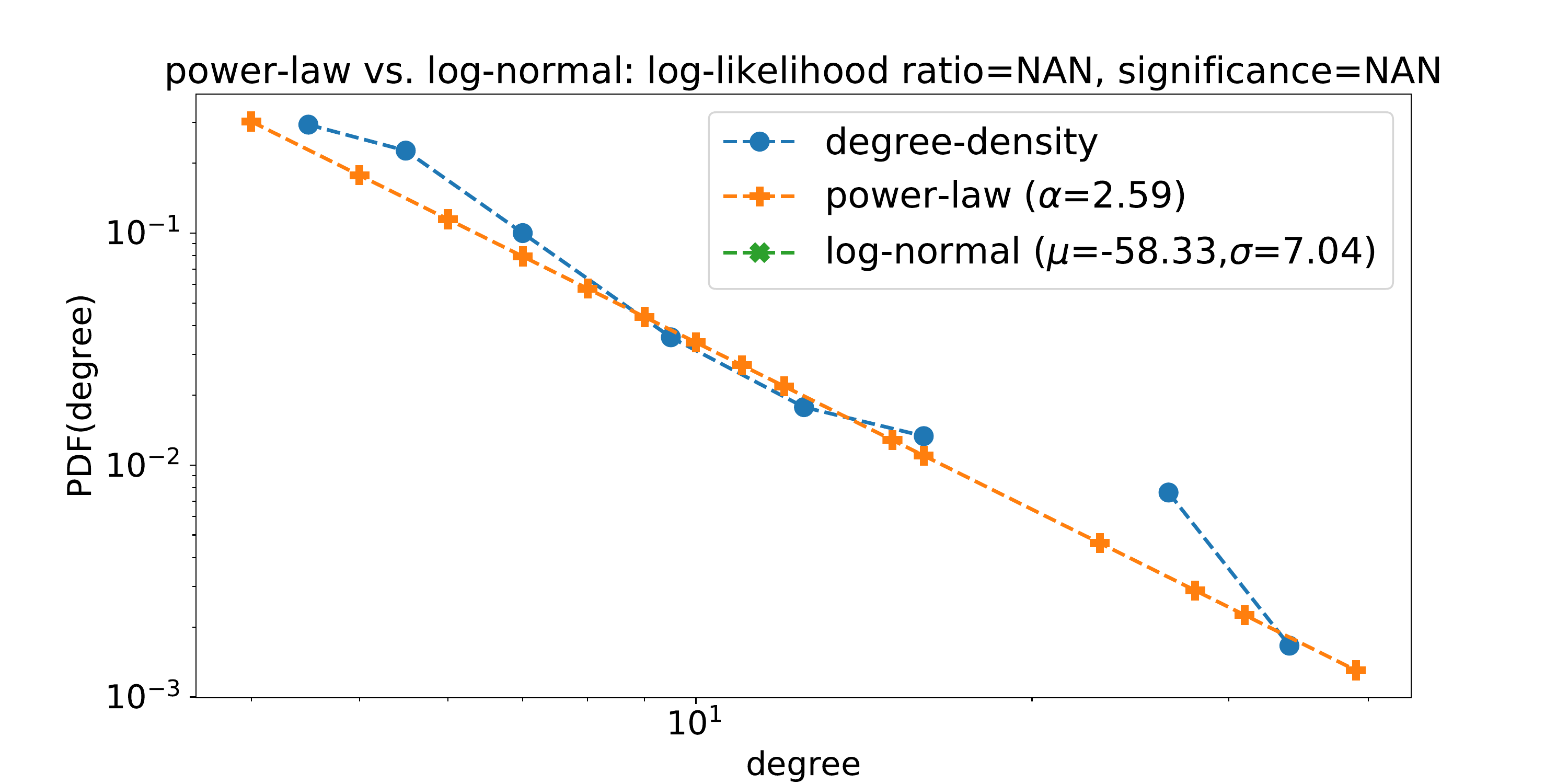}
        \caption{\usnpa.}
	    \label{fig:usg1_degree}
    \end{subfigure}
    \begin{subfigure}[b]{0.4\textwidth}
        \includegraphics[width=\textwidth]{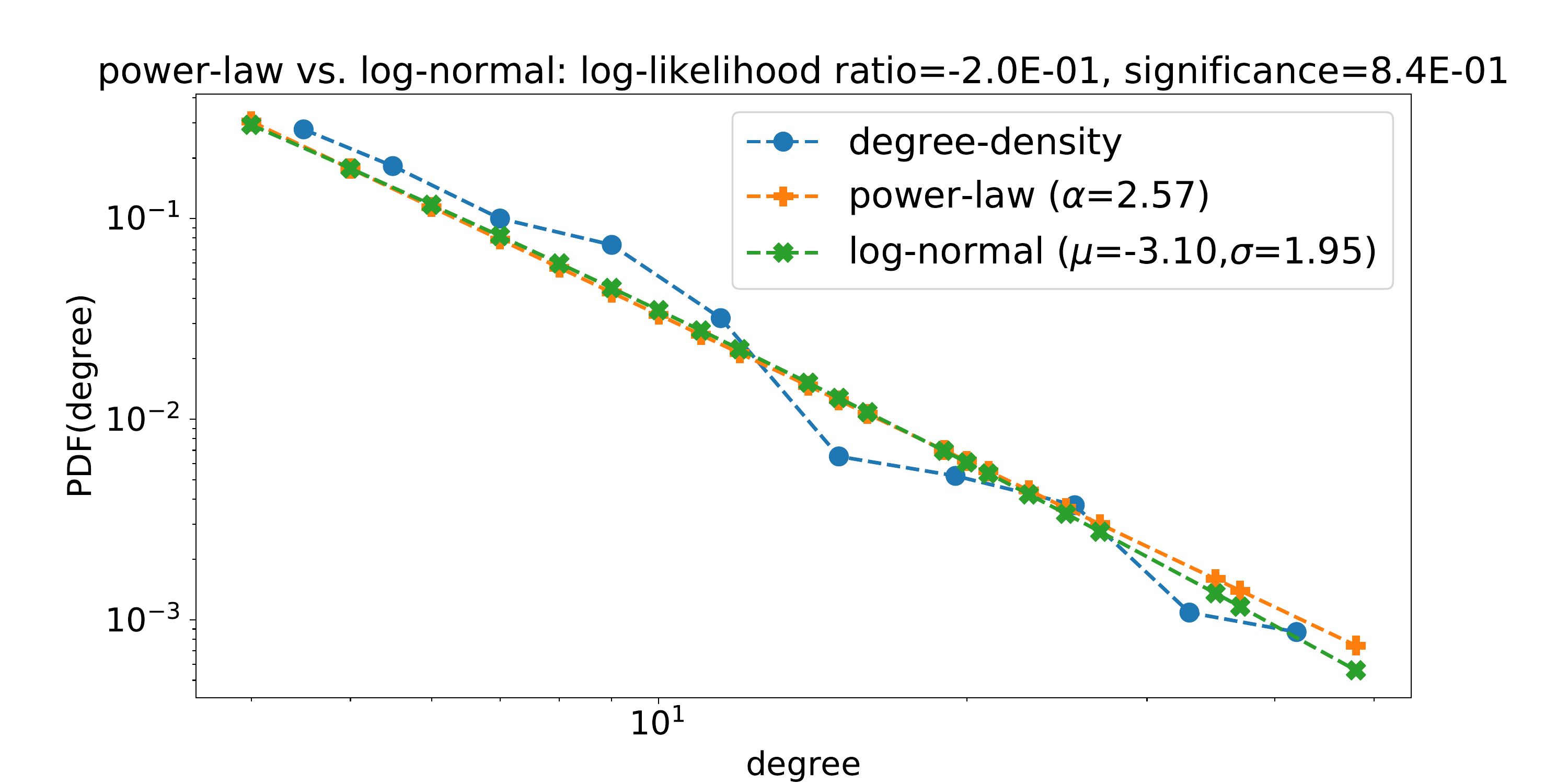}
        \caption{\usnpb.}
	    \label{fig:usg2_degree}
    \end{subfigure}
    
    \begin{subfigure}[b]{0.4\textwidth}
        \includegraphics[width=\textwidth]{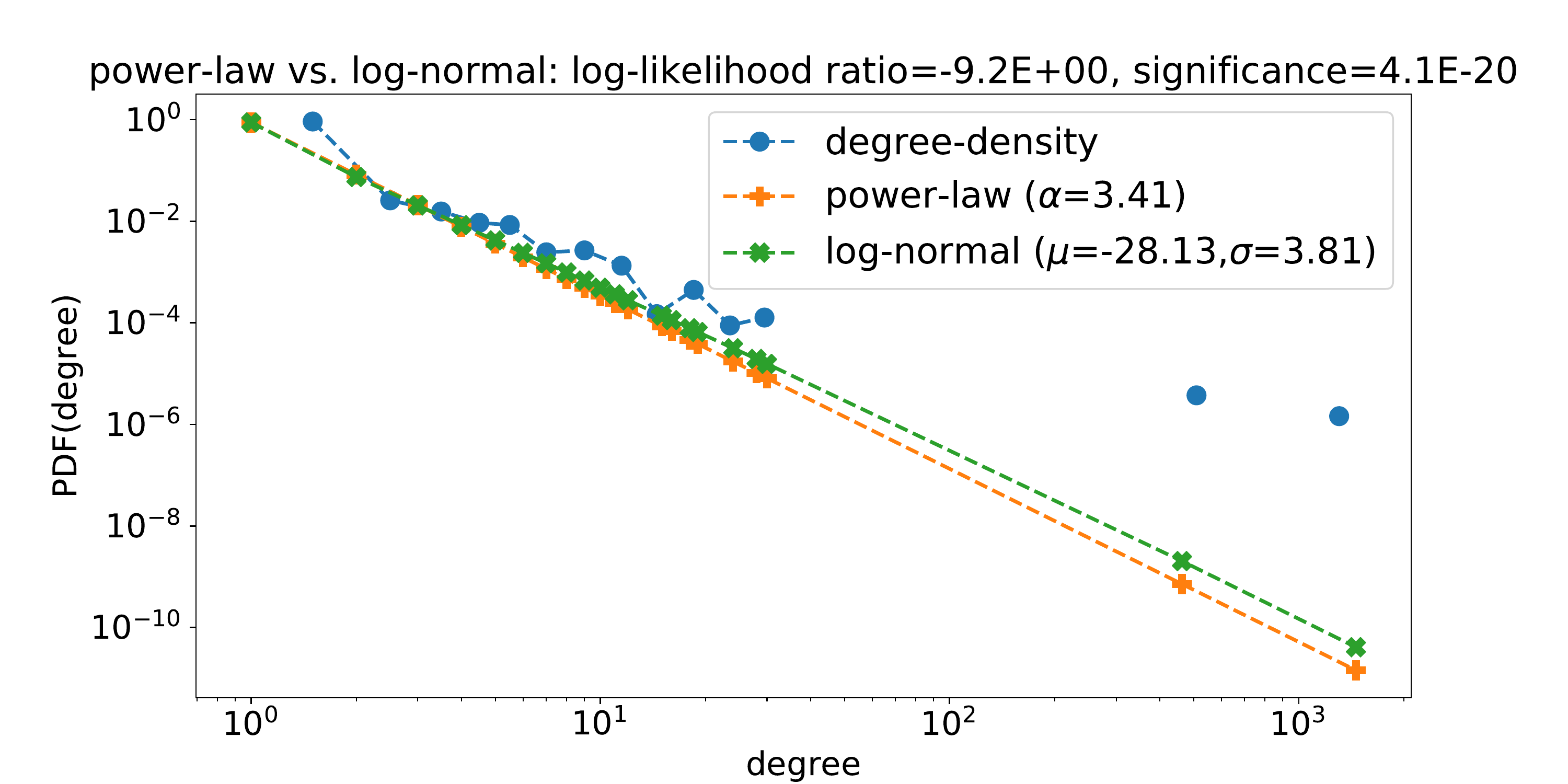}
        \caption{\usnpc.}
	    \label{fig:usg3_degree}
    \end{subfigure}
    
    \begin{subfigure}[b]{0.4\textwidth}
        \includegraphics[width=\textwidth]{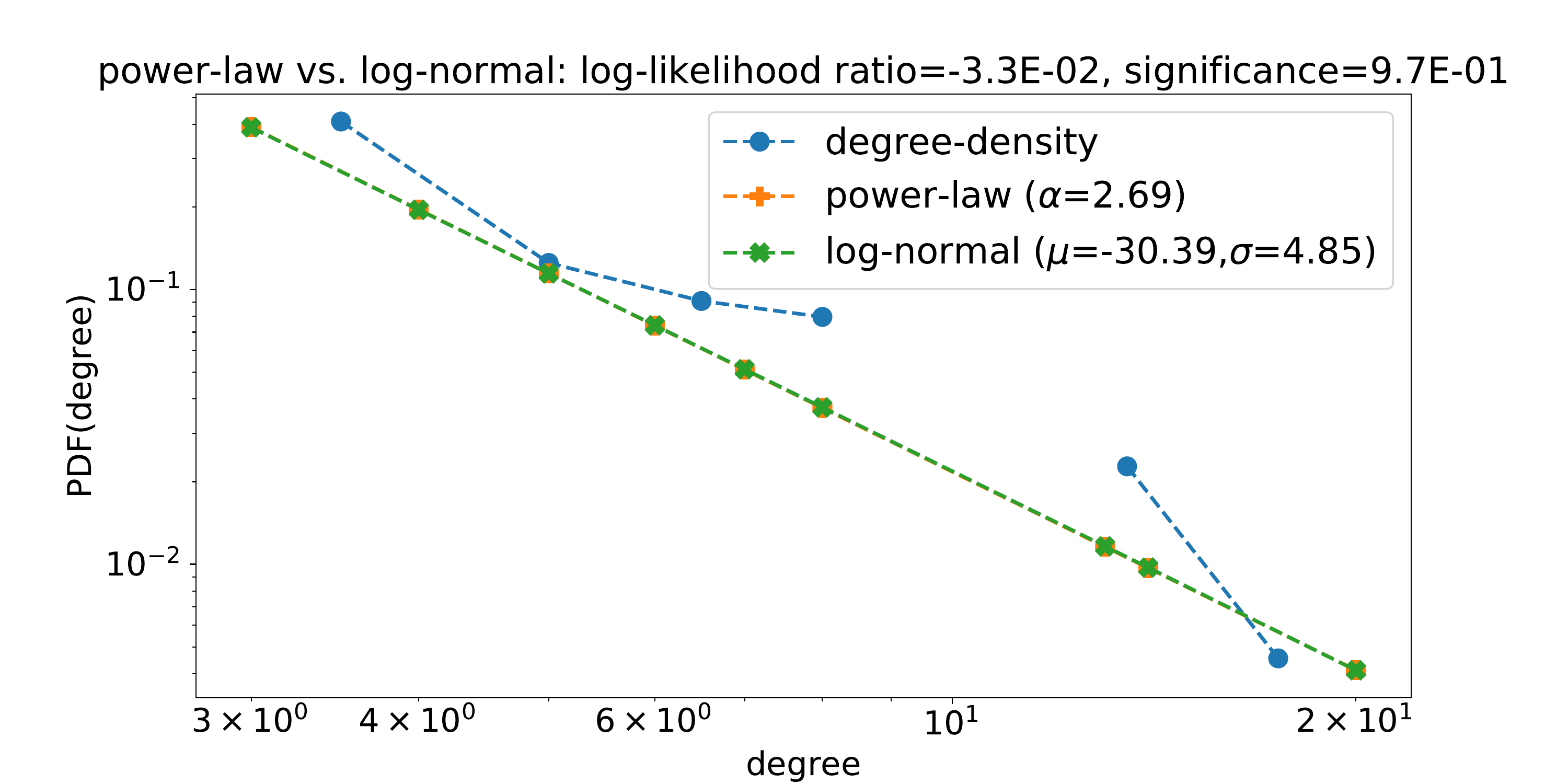}
        \caption{\ucore.}
	    \label{fig:usgi_degree}
    \end{subfigure}
    \begin{subfigure}[b]{0.4\textwidth}
        \includegraphics[width=\textwidth]{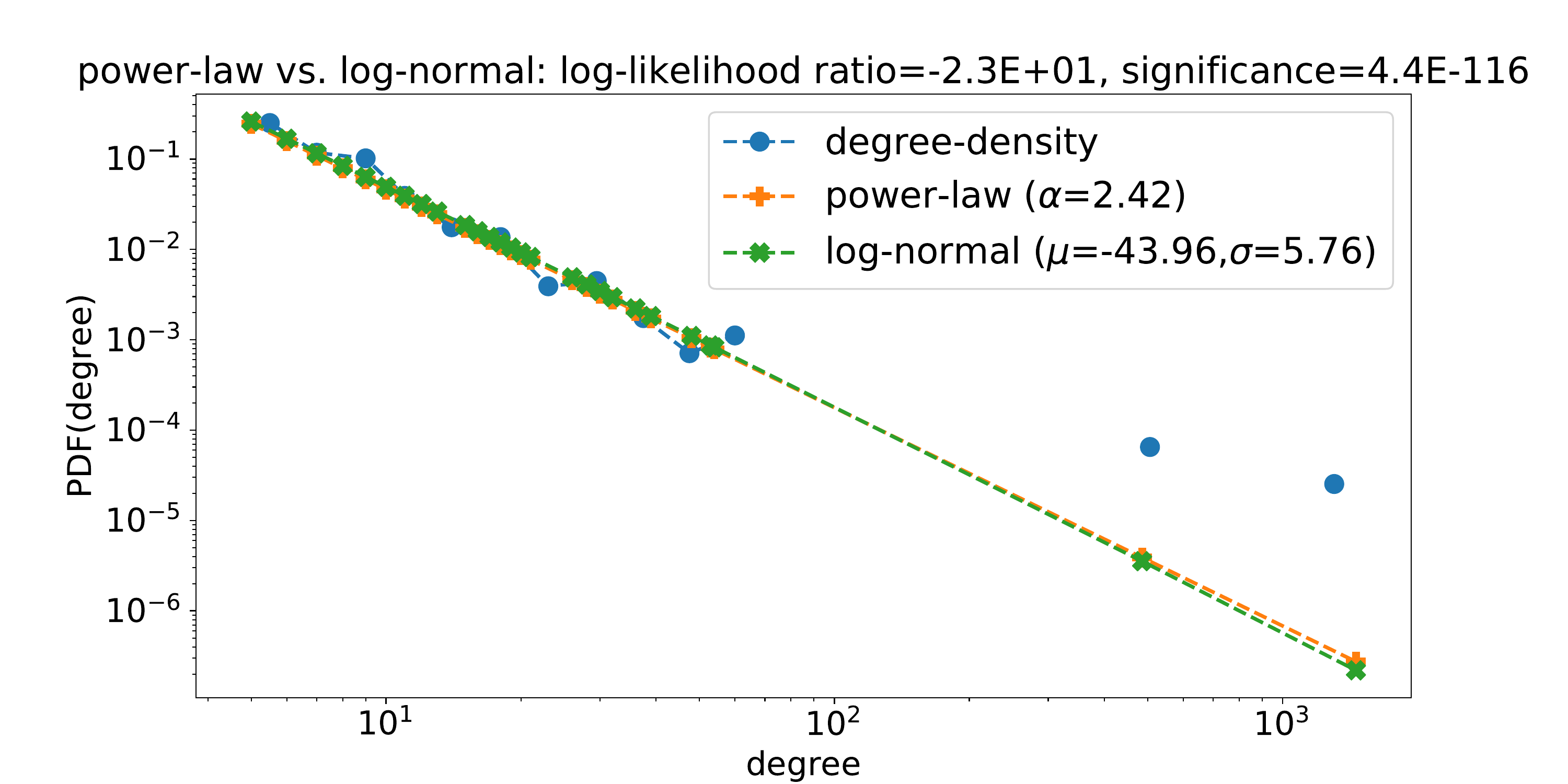}
        \caption{\uunion.}
	    \label{fig:usgu_degree}
    \end{subfigure}
    \caption{The degree distribution for the USGs.}
    \label{fig:degree_undirected}
\end{figure*}

In Figure~\ref{fig:degree_undirected}, we report the fitted degree distribution for the USGs.
These plots seem to broadly confirm the insights provided by the DSGs.
In addition, they show that only considering mutual connections is apparently an effective way for preserving the social structure of the graph when switching from a directed to an undirected graph.
Furthermore, apart from a couple of outliers in the union graph, the power-law fit is especially accurate in the two largest graphs -- the USG2 and the USGU -- somehow speaking in favour of the importance of keeping all available information together to capture the social dynamics of Tor.

\begin{figure*}[htbp]
    \centering
    \begin{subfigure}[b]{0.45\textwidth}
        \includegraphics[width=\textwidth]{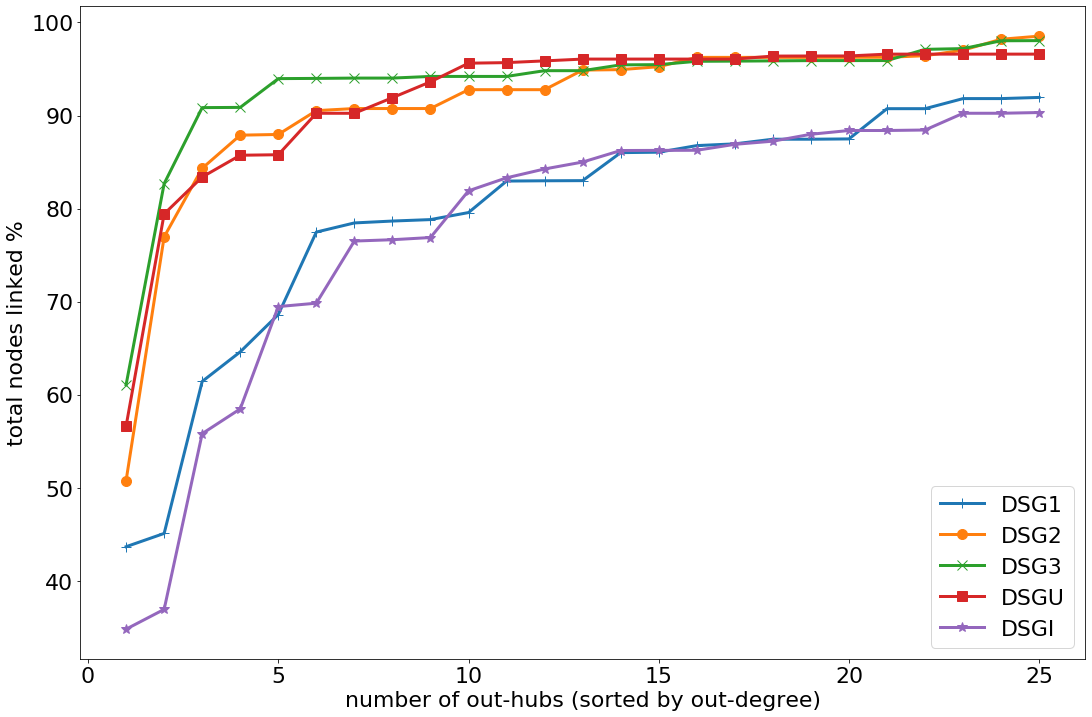}
        \caption{DSGs.}
	    \label{fig:dsg_cumulative}
    \end{subfigure}
    \hfill
    \begin{subfigure}[b]{0.45\textwidth}
        \includegraphics[width=\textwidth]{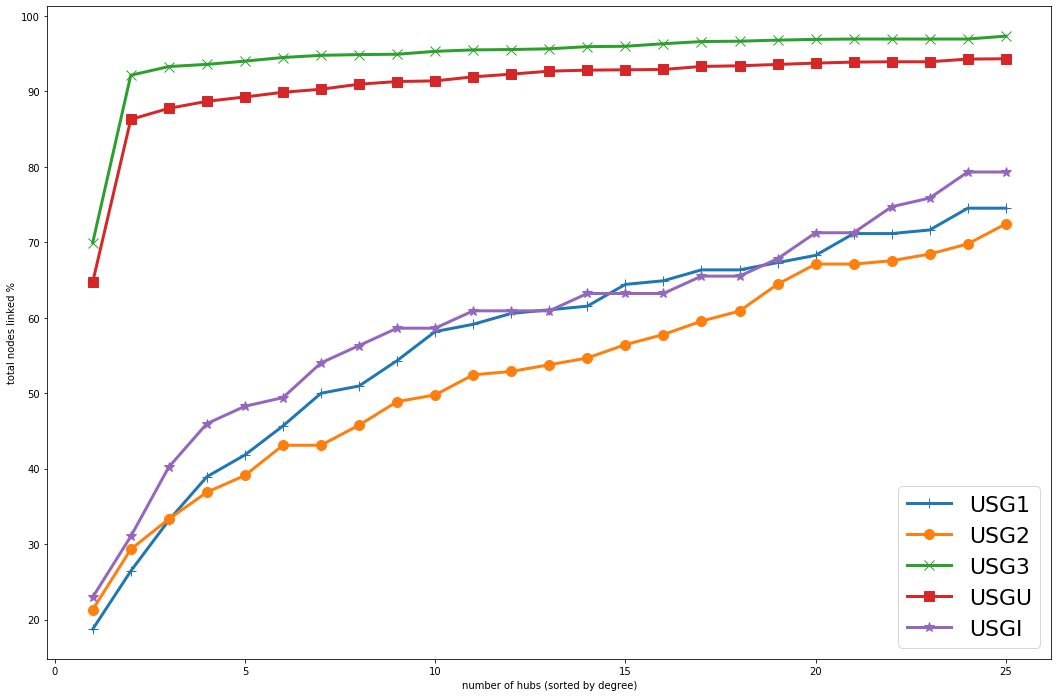}
        \caption{USGs.}
	    \label{fig:usg_cumulative}
    \end{subfigure}
    \caption{Cumulative percentage of the graph linked by the top hubs.}
    \label{fig:topout}
\end{figure*}

Motivated by the long tail of the degree distributions and with the purpose of gaining a better understanding of how the whole graph can be explored from just a few starting points, in Figure~\ref{fig:topout} we plot the cumulative percentage of the giant component of the network that is at distance one from the 25 top hubs -- \emph{i.e.}, top out-degree vertices in the DSGs and top degree vertices in the USGs.
We see that just a few out-hubs suffice to reach $\approx90\%$ of the graph\footnote{In DSGs, the giant component is \emph{de facto} the whole graph} in just one click in \dsnpb, \dsnpc as well as in \dunion.
In \dsnpa, however, we need the top-20 out-degree hidden services to reach the same percentage of the graph, albeit the top-6 out-degree services reach out to almost 80\% of the nodes.
This difference between \dsnpa and the other snapshots is reflected in the behavior of \dcore.
If we switch to the USGs, we see a completely different scenario.
The giant components of \usnpc and \uunion are dominated by just two services, with all other giving a negligible contribution.
On the other hand, the giant components of \usnpa, \usnpb and \ucore are much less centralized, in accordance with the much lower value for $\mathrm{Cen}$ reported in Table~\ref{tab:global_undirected}.
What is especially surprising by comparing figures~\ref{fig:dsg_cumulative} and~\ref{fig:usg_cumulative} is that shifting from directed to mutual connections seems to completely alter the topology of \dsnpb.
\subsection{Community Structure}\label{sec:clustering}
Let us now examine the community structure of the 10 Tor Web graphs.
We used the well-known Louvain algorithm~\cite{blondel2008fast}, based on modularity maximization.
As often done in the literature~\cite{girvan2002community}, we considered edge weights to make it harder to break an edge corresponding to a hyperlink that appears several times in the dataset.

\begin{figure*}[htbp]
    \centering
    \begin{subfigure}[b]{0.48\textwidth}
        \includegraphics[width=.8\textwidth]{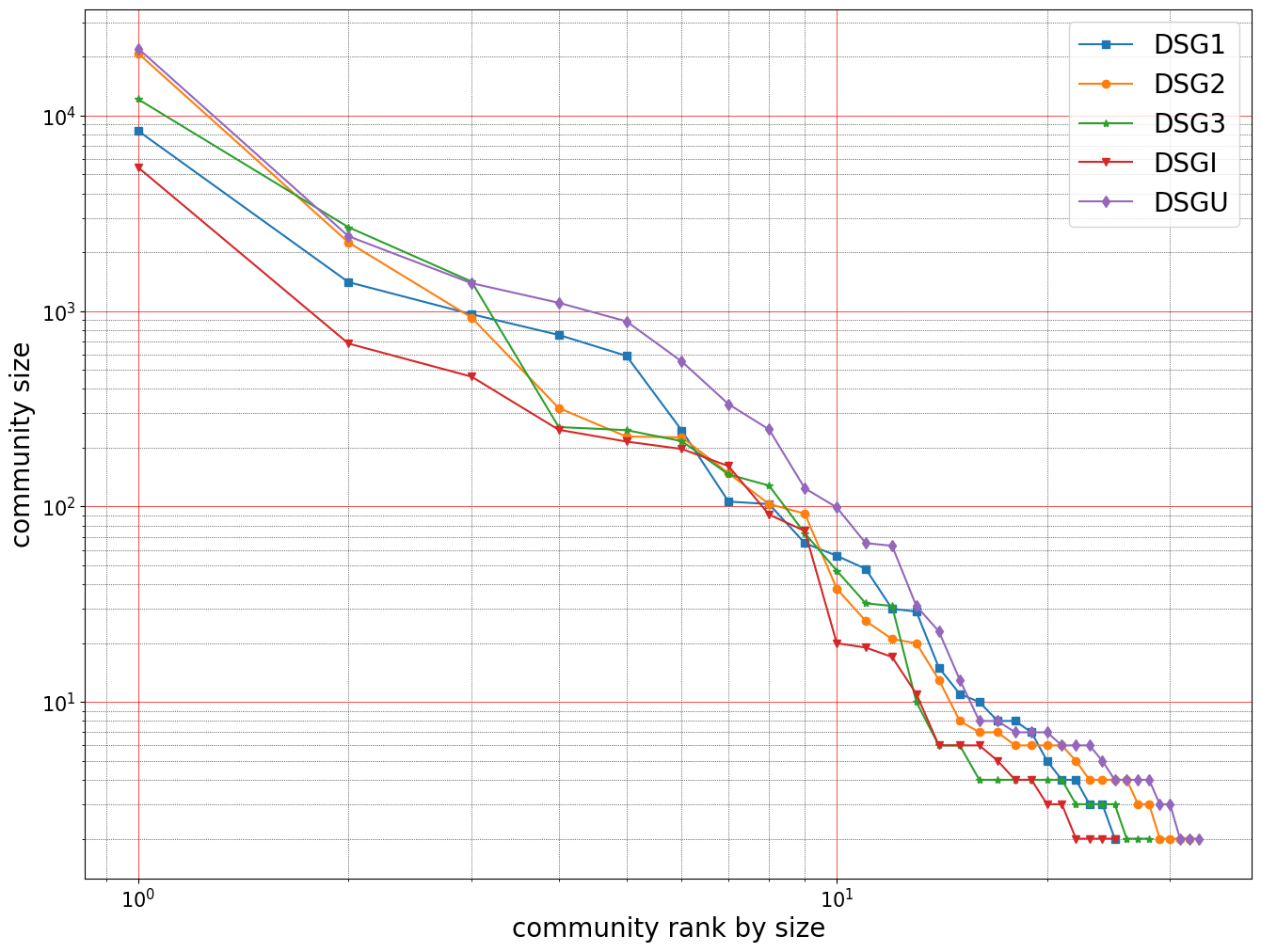}
        \caption{DSGs.}
        \label{fig:clustering_sizes_directed}
    \end{subfigure}
    \hfill
    \begin{subfigure}[b]{0.48\textwidth}
        \includegraphics[width=.8\textwidth]{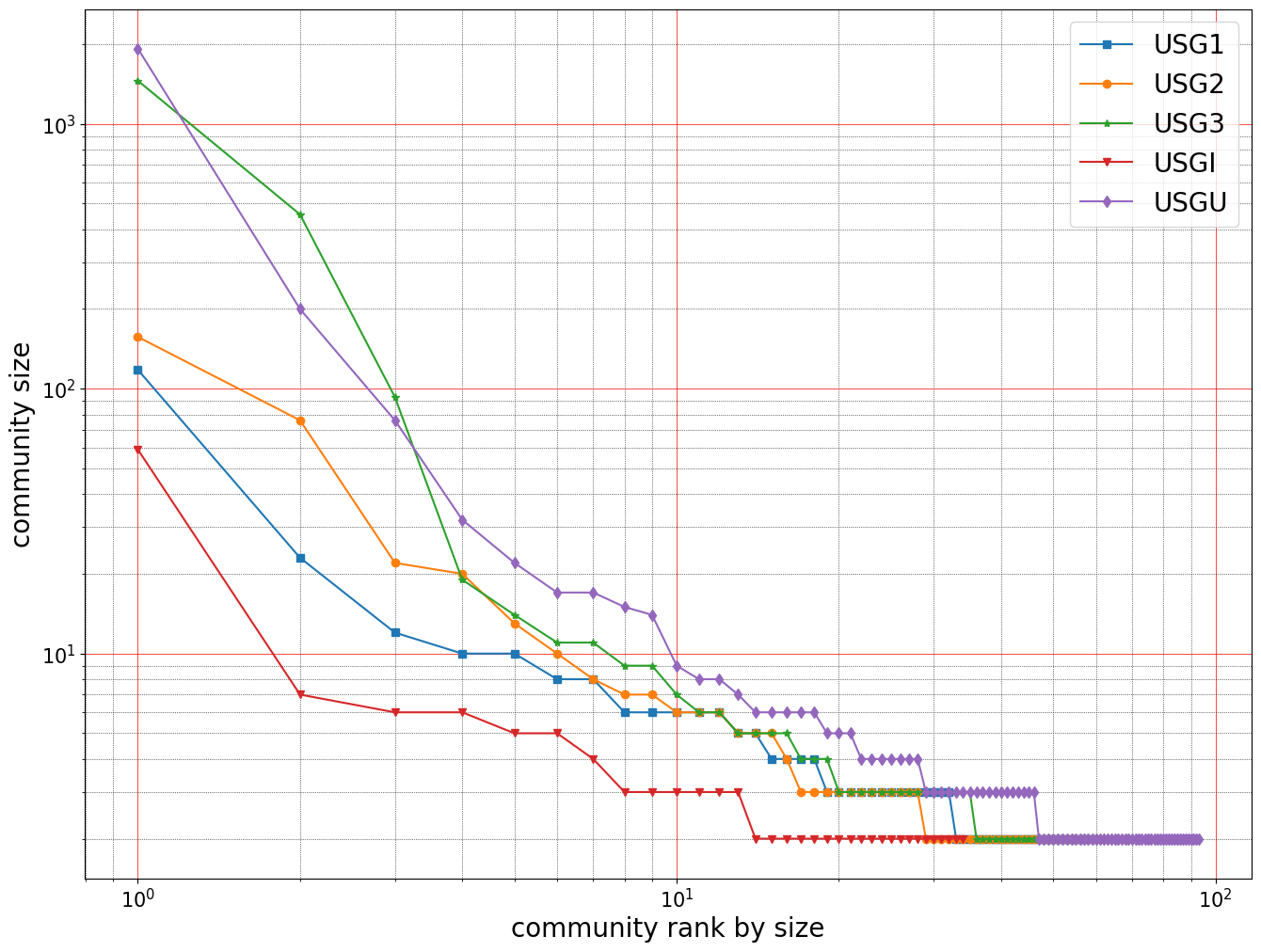}
        \caption{USGs.}
        \label{fig:clustering_sizes_undirected}
    \end{subfigure}
    
    \caption{The community size distribution for our Tor Web graphs.}
    \label{fig:clustering_sizes}
\end{figure*}

\begin{figure*}[htbp]
    \centering
    \begin{subfigure}[b]{0.49\textwidth}
        \includegraphics[width=.8\textwidth]{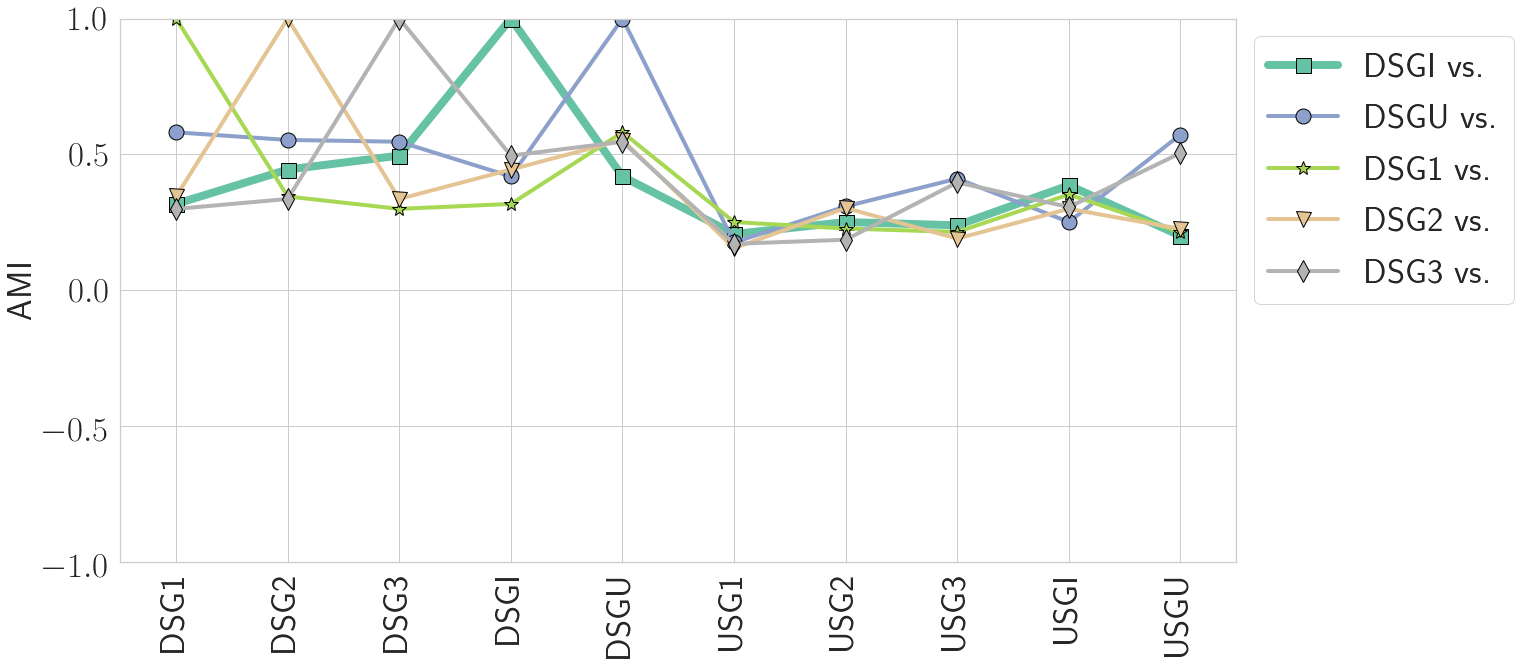}
        \caption{DSGs \emph{vs.} all graphs.}
        \label{fig:clustering_distance_directed}
    \end{subfigure}
    \begin{subfigure}[b]{0.49\textwidth}
        \includegraphics[width=.8\textwidth]{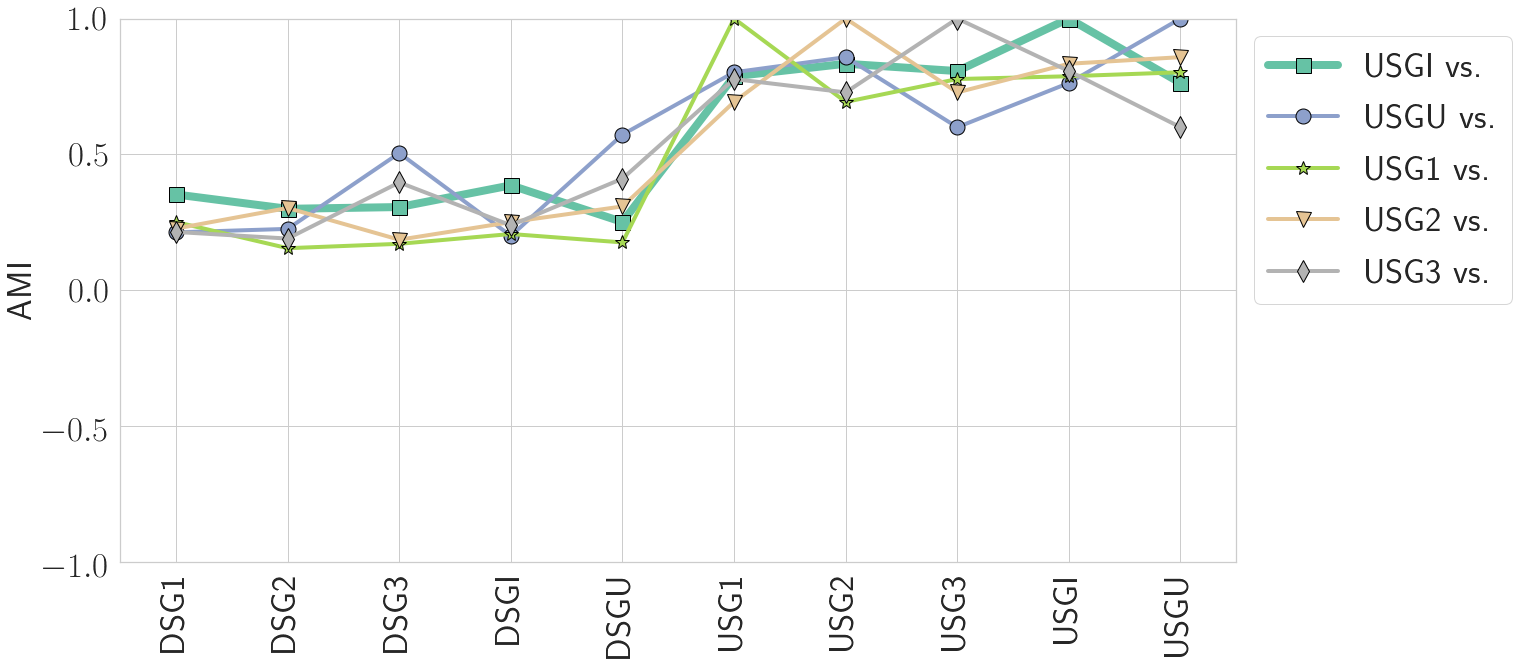}
        \caption{USGs \emph{vs.} all graphs.}
        \label{fig:clustering_distance_undirected}
    \end{subfigure}
    
    \caption{The comparison of the partitions obtained for our Tor Web graphs.}
    \label{fig:clustering_comparison}
\end{figure*}

In Figures~\ref{fig:clustering_sizes} and~\ref{fig:clustering_comparison} we compare the obtained community structures for all of our graphs.
First, in Figures~\ref{fig:clustering_sizes} we plot the distribution of cluster sizes for the DSGs and USGs, respectively.
These two plots highlight that, at a high level, all DSGs appear to have a very similar structure, in terms of number and size of the clusters.
This sort of similarity is still partially visible in the USGs, with the significant difference in the graph size apparently mostly impacting the size of the greatest communities.
It is possibly more important to assess to which extent the obtained clusters are influenced by the network's volatility or, in other words, whether they are coherent across different graphs -- an element in favor of the possibility that pinpointed clusters have an intrinsic meaning.
In Figure~\ref{fig:clustering_comparison} we use the well-known Adjusted Mutual Information (AMI) to compare the clusters emerged across different graphs.
We recall that the AMI of two partitions is 1 if the two partitions are identical, it is 0 if the mutual information of the two partitions is the expected mutual information of two random partitions\footnote{Here, the meaning of ``random'' depends on the choice of a distribution over the set of all possible partitions~\cite{vinh2009information}}, and it is negative if the mutual information of the two partitions is worse than the expected one.
We see that common vertices are clustered in an extremely stable way in the USGs, meaning that the existence of a mutual link is -- as expected -- a stronger indicator of the similarity between two services.
We also see that the union graphs \dunion and \uunion, \emph{i.e.}, the graphs based on all collected data, are those whose community structure is less influenced by switching from the directed to the undirected graph.
In some sense, this means that the clustering obtained for \dunion can be reasonably considered as an extension of the very meaningful partition obtained for \uunion. 
\subsection{Bow-Tie Structure}\label{sec:bowtie}
Finally, as commonly done to describe \web graphs~\cite{broder2000graph}, in Table~\ref{tab:bowtie} we provide a bow-tie decomposition of our directed graphs, compared with previous work.
Our findings broadly confirm what emerged in~\cite{griffith2017graph}, \emph{i.e.}, that in the \tor Web the LSSC is very small and everything else falls into the OUT component -- giving rise to a significantly different structure with respect to the WWW.
However, by comparing three large crawls of the \tor Web we highlight a few features that were not noticed before.
On the one hand, the share of the LSCC in the total size of the graph is very variable over time, to the point that in \dsnpc it is $\sim4\times$ larger than in the other two snapshots.
On the other hand, the structure of the \dcore graph is slightly more similar to the WWW, with all components being non-empty.
\begin{table*}[ht!]
\centering
\begin{threeparttable}
\caption{Bow-Tie structure}
\begin{tabular}{lrrrrrr}
\toprule
 \multirow{2}{*}{Graph} &  \multicolumn{6}{c}{Component} \\
                        &  LSCC   &    IN    &   OUT   &   TUBES & TENDRILS & DISCONNECTED \\
\midrule
\rowcolor{gray!10}      & 22.3M   &  3.3M    &  13.3M  &    17K  &    514K  &  3.5M        \\
\rowcolor{gray!10}  \multirow{-2}{*}{WWW from~\cite{Lehmberg:2014:GSW:2615569.2615674}}
                        & 51.94\% & 7.65\%   & 30.98\% & 0.04\%  & 1.2\%    & 8.2\%        \\
\multirow{2}{*}{\tor from~\cite{griffith2017graph}}
                        &  297    &   0      &   6881  & 0       & 0       &  0            \\
                        &  4.14\% &  0.0\%   & 95.86\% & 0.0\%   & 0.0\%   &  0.0\%        \\
\rowcolor{gray!10}      &   466   &  0       & 12363   & 0       & 0       & 0             \\
\rowcolor{gray!10} \multirow{-2}{*}{\dsnpa}
                        &  3.63\% &  0.0\%   & 96.37\% & 0.0\%   & 0.0\%   & 0.0\%         \\
\multirow{2}{*}{\dsnpb}
                        & 820     & 0        & 24488   &  0      & 0       & 0             \\
                        & 3.24\%  & 0.0\%    & 96.76\% &  0.0\%  & 0.0\%   & 0.0\%         \\
\rowcolor{gray!10}      & 2371    &  0       & 15089   &  0      & 0       & 0             \\
\rowcolor{gray!10} \multirow{-2}{*}{\dsnpc}
                        & 13.58\% & 0.0\%    & 86.42\% &  0.0\%  & 0.0\%   & 0.0\%         \\
\multirow{2}{*}{\dcore}
                        & 169     & 9        & 7415    &  1      & 74      & 1             \\
                        & 2.2\%   & 0.12\%   & 96.69\% &  0.01\% & 0.97\%  & 0.01\%        \\
\rowcolor{gray!10}      & 3062    & 0        & 26411   & 0       & 0       & 0             \\
\rowcolor{gray!10} \multirow{-2}{*}{\dunion}
                        & 10.39\% & 0.0\%    & 89.61\% &  0.0\%  &  0.0\%  &  0.0\%        \\
\bottomrule
\end{tabular}
\begin{tablenotes}[flushleft]
\scriptsize
\item LSCC is the largest strongly connected component.
\item IN is the set of nodes $v\in V\setminus \mathrm{LSCC}$ such that there is a path from $v$ to LSCC. 
\item OUT is the set of nodes $v\in V\setminus \mathrm{LSCC}$ such that there is a path from LSCC to $v$. 
\item TUBES is the set of nodes $v\in V\setminus \left(\mathrm{LSCC}\cup\mathrm{IN}\cup\mathrm{OUT}\right)$ such that there is a path from IN to $v$ as well as a path from $v$ to OUT. 
\item TENDRILS is the set of nodes $v\in V\setminus \left(\mathrm{LSCC}\cup\mathrm{IN}\cup\mathrm{OUT}\right)$ such that there is either a path from IN to $v$ or a path from $v$ to OUT, but not both. 
\item DISCONNECTED is the set of all other nodes $v\in V\setminus \left(\mathrm{LSCC}\cup\mathrm{IN}\cup\mathrm{OUT}\cup\mathrm{TUBES}\cup\mathrm{TENDRILS}\right)$.
\end{tablenotes}
\label{tab:bowtie}
\end{threeparttable}
\end{table*}
%
\subsection{Discussion}
Summarizing our results, we conclude that, albeit the \tor web graph presents a few common features of other real world networks, it has a significantly different structure with respect to the WWW.
In the \tor \web the LSSC is very small and everything else falls into the OUT component.
Only the structure of the \dcore graph is slightly more similar to the WWW, with all components being non-empty.

Generally speaking, \tor is a \emph{small world} network composed by a large percentage of volatile hidden services.
The network is characterized by the presence of in and out-hubs nodes that are critical for the graph's connectivity.
The hubs are persistent/stable nodes whose degree is heavily influenced by non-persistent nodes. 
In particular, there are few in-hubs and several large out-hubs in the network -- \emph{i.e.}, link directories or similar web services.
More than half of the network's nodes are reachable in just one click from the top out-hubs.
Peripheral services are however loosely connected, making the \tor \web an inefficient network.

Although many \tor nodes are not persistent, the graph on the whole seems to possess a meaningful and stable community structure.
This is especially visible when only mutual connections are considered.
``Bidirectional'' ties not only provide more consistent communities, but they generally better preserve a few social structural features of the graph, such as the tendency to cluster of nodes having degree in some intermediate range.
However, bidirectional paths in the \tor network are generally long -- when they do exist.
Considering only mutual connections, the Tor web is thus not a \emph{small world} anymore.

Global metrics of the graph are widely consistent among snapshots.
This means that the volatility of peripheral nodes does not heavily influence the global structure of the network.
The \dunion and \dcore seem to capture all the features of the snapshots, reflecting in different ways some specific features that appear only in few of them. For these reasons in the next Sections we use only the intersection and union graphs to avoid the drawbacks of \emph{information overload}.

\section{Characterizing \tor Services}\label{sec:local}

Hereafter, we shift the focus onto vertex properties of the Tor Web graph.
With the purpose of sorting out the possible roles of a service in the network, we first provide a correlation analysis of several local structural properties.
Among the considered set of metrics -- recapped in Table~\ref{tab:local_metrics} -- we include the \emph{hubscore} and \emph{authscore} provided by the HITS algorithm~\cite{kleinberg1999authoritative}.
These two metrics allow to identify and characterize the out-hubs (``hubs'', \emph{i.e.}, high hubscore) and in-hubs (``authorities'', \emph{i.e.}, high authscore) identified in Section~\ref{sec:topology}, in a more proper yet conceptually similar way. 
We then compare semantic and topological features by making use of the DUTA~\cite{al2019torank} dataset of manually labeled hidden services.
We measure the similarity of content-based and modularity-based clusterings and we determine to which extent a service's properties are related to its contents/structure.
As already discussed in Section~\ref{sec:topology} we consider only the intersection and union graphs for the sake of clarity.

\begin{table}[htbp]
\caption{Local metrics notations and definitions.}
\label{tab:local_metrics}
\centering
\begin{tabular}{ll}
\toprule
\textbf{Short name} & \textbf{Full name and definition\footnotemark} \\
\midrule
\BC           & Betweenness centrality: $BC(v)=\sum_{s\neq v \neq t\in V} \frac{\sigma_{st}(v)}{\sigma_{st}}$ \\		    
\CC           & Closeness centrality: $CC(v) = \frac{N-1}{\sum_{u\in V} d(u,v)}$ \\
\PR           & PageRank: see~\cite{franceschet2011pagerank} \\
\auth         & Authority score: see~\cite{kleinberg1999the} \\
\hub          & Hub score: see~\cite{kleinberg1999the} \\
\eff          & Efficiency: $E(v) = \frac{1}{deg(v)(deg(v)-1)}\sum_{u\neq w:\; v \to u \wedge v \to w} \frac{1}{d(u,w)}$ \\
\tran             & Transitivity: $T(v) = \frac{\# (u,w):\; v \to u \wedge v \to w \wedge (u \to w \vee w \to u)}{\# (u,w):\; v\to u \wedge v \to w}$ \\ 
\ecc      & Eccentricity: $\epsilon(v) = \max_{u\in V} d(v,u)$ \\
\lcratio       & Links-to-chars ratio, see Section~\ref{sec:dataset} \\
\bottomrule
\end{tabular}
\end{table}
\footnotetext{Beware that some of these metrics are only defined for directed graphs.}

\subsection{Correlation Analysis}\label{sec:correlation}

\begin{figure}[htbp]
    \centering
    \begin{subfigure}[b]{0.45\textwidth}
        \includegraphics[width=\textwidth]{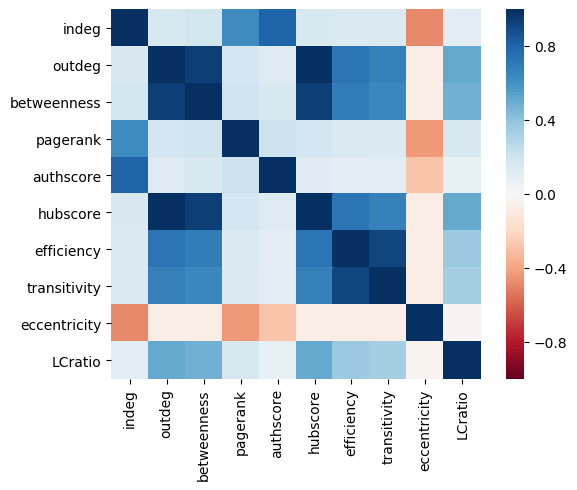}
        \caption{\dcore.}
	    \label{fig:dcore_correlation}
    \end{subfigure}
    \begin{subfigure}[b]{0.45\textwidth}
        \includegraphics[width=\textwidth]{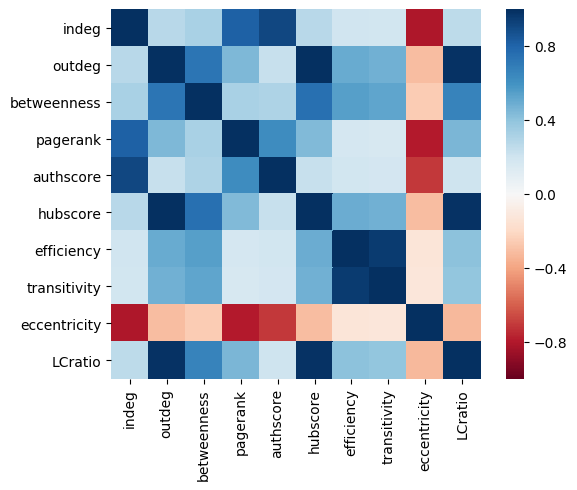}
        \caption{\dunion.}
	    \label{fig:dunion_correlation}
    \end{subfigure}
    
    \begin{subfigure}[b]{0.45\textwidth}
        \includegraphics[width=\textwidth]{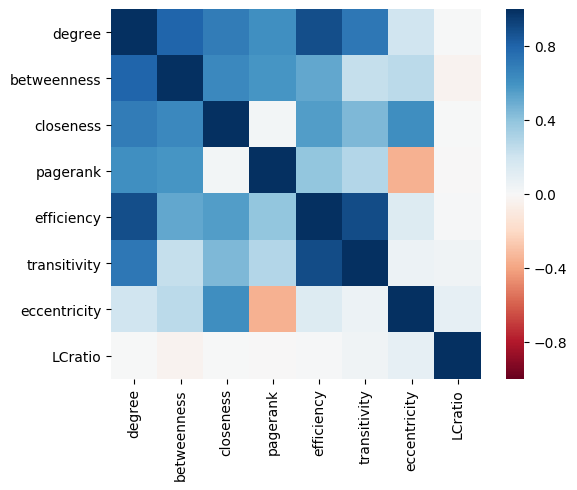}
        \caption{\ucore.}
	    \label{fig:ucore_correlation}
    \end{subfigure}
    \begin{subfigure}[b]{0.45\textwidth}
        \includegraphics[width=\textwidth]{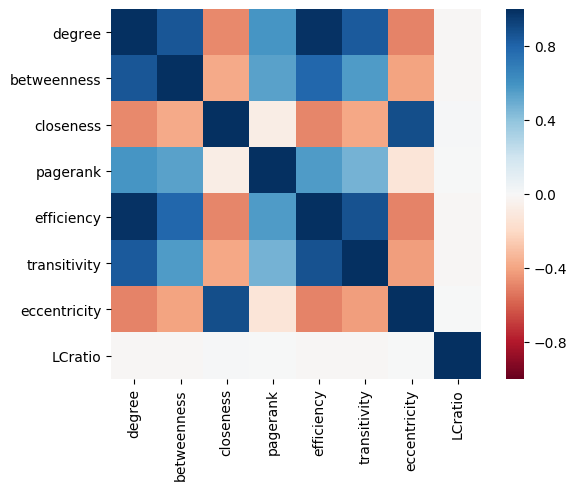}
        \caption{\uunion.}
	    \label{fig:uunion_correlation}
    \end{subfigure}
    \caption{Spearman's rank correlation coefficient between the considered local metrics.}
    \label{fig:correlation}
\end{figure}

Figure~\ref{fig:correlation} visually shows the pairwise correlation of the metrics defined in Table~\ref{tab:local_metrics}.
We rely on Spearman's rank correlation coefficient -- rather than the widely used Pearson's -- for a number of reasons: (i) we are neither especially interested in verifying linear dependence, nor we do expect to find it; (ii) we argue that not all the considered metrics yield a clearly defined interval scale -- while they evidently provide a ordinal scale; (iii) when either of the two distributions of interest has a long tail, Spearman's is usually preferable because the rank transformation compensates for asymmetries in the data; and (iv) recent work~\cite{litvak2013uncovering} showed that Pearson’s may have pathological behaviors in large scale-free networks.

In the DSGs (figures~\ref{fig:dcore_correlation} and~\ref{fig:dunion_correlation}) we notice a few interesting trends -- albeit not entirely surprising.
The authscore tends to correlate more with the in-degree, closeness and pagerank, whereas the hubscore tends to correlate more with the out-degree, betweenness, efficiency, transitivity and \lcratio.
In other words, vertices that are authoritative are, on average, easier to reach and may not be hubs.
Hubs, on the other hand, are not necessarily authoritative, they facilitate information flows and are at the center of highly clustered regions.
The \lcratio  seems to perform pretty well, on average, as a measure of \emph{hubbiness}, as expected.
The eccentricity is instead uncorrelated or negatively correlated with all other metrics.
This says that central nodes are either close to or entirely disconnected to any other service, while long paths exist that connect peripheral services.

The most remarkable aspect emerging from the correlation analysis of the USGs is probably the great impact that switching to mutual connections has on the distribution of local metrics.
The results for \ucore and \uunion are not only very different from their directed counterparts, but they also significantly differ from each other.
In the \ucore what stands out is the lack of correlation between closeness and pagerank, and between the \lcratio and all other metrics.
In the \uunion we instead notice a very interesting phenomenon: the closeness and the eccentricity ``agree'' with each other while they negatively correlate with all other measures.
%
\subsection{Content-based Classification of Services}\label{sec:DUTA}
\begin{figure*}[htbp]
 \centering
        \includegraphics[width=.88\textwidth]{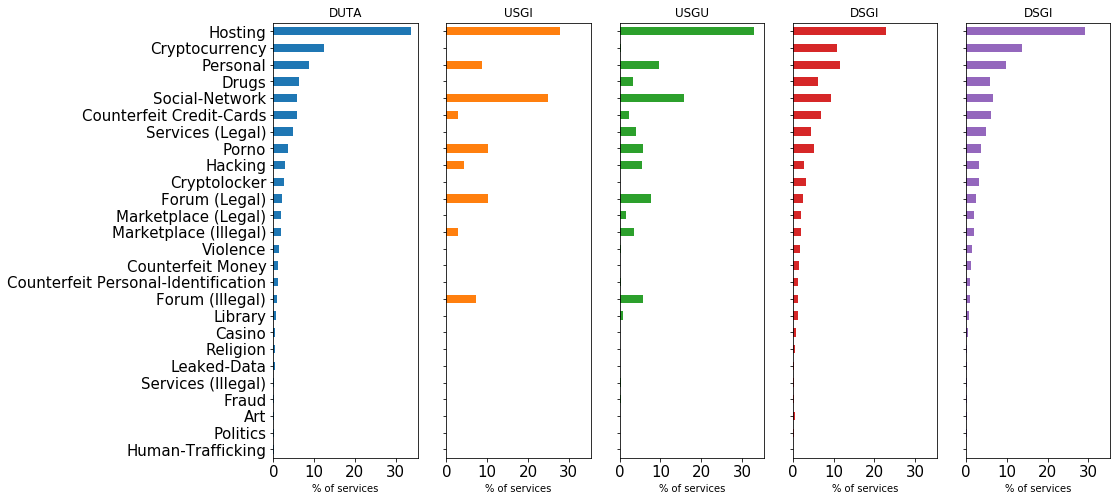}
        \caption{Distribution of tags from Table~\ref{tab:duta} in the DUTA dataset and in the four considered \tor \web graphs.}
    \label{fig:duta_distribution}
\end{figure*}

\begin{figure*}[htbp]
 \centering
        \includegraphics[width=.65\textwidth]{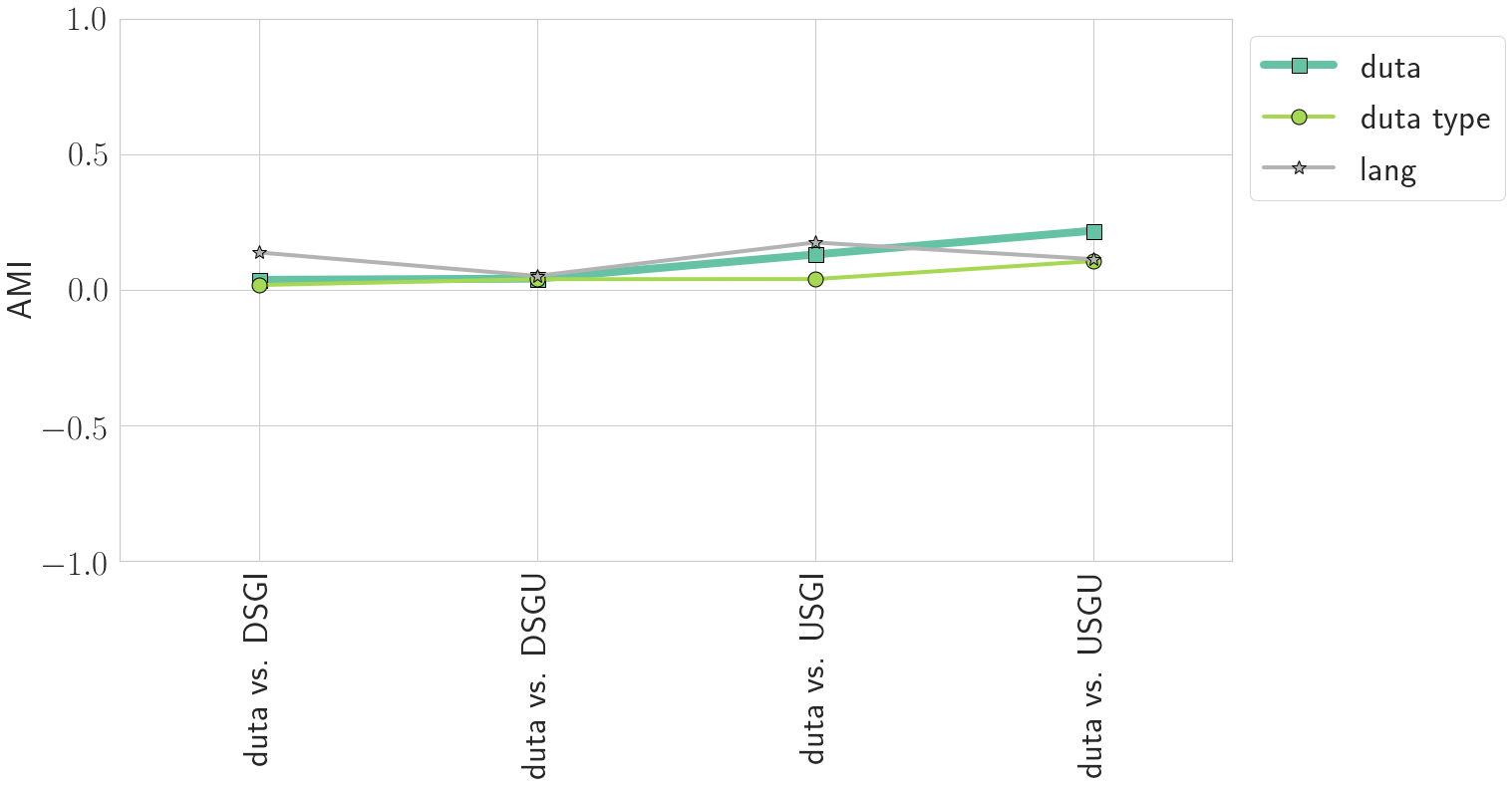}
        \caption{The comparison of the topic-based partition induced by the DUTA dataset and the modularity-based partitions obtained through Louvain's algorithm on our graphs.}
    \label{fig:duta_AMI}
\end{figure*}

For contents analysis we rely on the DUTA dataset, the widest publicly available thematic dataset for Tor, consisting of a three-layer classification of 10250 hidden services~\cite{al2017classifying,al2019torank}.
Albeit the DUTA dataset does not cover our graphs entirely, it has the undeniable advantage of being manually tagged -- by choosing it rather than carrying out a fresh new classification of our dataset, we trade coverage for accuracy.
The percentage of vertices of our graphs contained in the DUTA dataset is significant, especially for the two intersection graphs ($49.5\%$ and $96\%$, respectively, dropping to $28\%$ and $24.3\%$ for the unions). 
Further, if we only consider the first $200$ nodes ordered by out-degree, only $\approx15\%$ are not tagged.
The DUTA dataset provides a two-layers thematic classification plus a language tag for each service.
The thematic classes are further categorized as ``Normal'',   ``Suspicious'' or ``Unknown''.
The ``Unknown'' category only includes classes that correspond to services whose nature was impossible to establish: ``Empty'', ``Locked'' or ``Down''.
Due to the limited information provided by these tags, we ignore all ``Unknown'' services in the following.
For certain first layer classes (\emph{e.g.}, ``Marketplace'') that may be both ``Suspicious'' and ``Normal'', the second layer is exactly used to tell apart ``Legal'' and ``Illegal'' content.
We consider the second layer for this purpose only, thus obtaining the customized version of the DUTA thematic classification reported in Table~\ref{tab:duta}.
\begin{table*}[htbp]
\caption{The content-based classification used in this paper.}
\label{tab:duta}
\centering
\begin{tabular}{p{\widthof{Counterfeit Personal-Identification}}p{\widthof{Counterfeit Personal-Identification}}}
\toprule
\multicolumn{2}{c}{\textbf{Class name by type}}\\
\multicolumn{1}{c}{\textbf{Normal}} & \multicolumn{1}{c}{\textbf{Suspicious}}\\
\midrule
Art & Counterfeit Credit-Cards \\
Casino & 							Counterfeit Money\\
Cryptocurrency & 					Counterfeit Personal-Identification\\
Forum (Legal) & 					Cryptolocker\\
Hosting & 							Drugs\\
Library & 							Forum (Illegal)\\
Marketplace (Legal) & 				Fraud\\
Personal & 						Hacking\\
Politics & 						Human-Trafficking\\
Religion & 						Leaked-Data\\
Services (Legal) & 				Marketplace (Illegal)\\
Social-Network & 					Porno \\
				&					Services (Illegal) \\
      			&					Violence \\
\bottomrule
\end{tabular}
\end{table*}
In Figure~\ref{fig:duta_distribution} we compare the distribution of thematic tags in the DUTA dataset and in the four intersection and union graphs.
We immediately see that ``Hosting'' services are predominant in all cases.
We also see that the distribution in both the DSGI and the DSGU follows the original distribution quite closely, suggesting that the volatility of \tor's hidden services is unrelated to their content.
Finally, we notice that in the USGI and in the USGU, instead, some common classes are entirely missing or barely present (\emph{e.g.}, ``Cryptocurrency'') while some others are relatively much more frequent than in DUTA (\emph{e.g.}, ``Social Network'').
It is interesting that the latter are mostly classes related to sociality in a broad sense, again corroborating the idea that mutual connections better capture the social structure of \tor.
Once we have assessed the global prevalence of different classes of services in our graphs, we aim at understanding whether a more specific pattern emerges when we focus on modularity-based communities.
Since a single label from Table~\ref{tab:duta} is assigned to each service, the DUTA classification naturally induces three hard partitions, denoted ``duta'' (the individual classes), ``duta type'' (the macro categories ``Normal'' and ``Suspicious'') and ``lang'' (the language) in the following.
For the set of hidden services the our graphs share with the DUTA dataset, we can assess the coherence of topic-based and modularity-based clustering by plotting the AMI of ``duta'', ``duta type'' and ``lang'' with respect to the Louvain's clusters discussed in Section~\ref{sec:clustering}.
From Figure~\ref{fig:duta_AMI} it emerges very clearly that modularity-based clusters are \emph{not} thematically uniform, since the mutual information of the two partitions is always barely greater than the mutual information of two random partitions.
Our analysis makes clear that DUTA clusters and Louvain's clusters are substantially unrelated.
%
\subsection{Topological Features for Content-based Classification}~\label{sec:features}

Finally, in this section we measure the information gain provided by topological vertex properties with respect to content-based classification.
To this end, we proceed as follows:
\begin{itemize}
    \item For each category $C$, we consider the dummy variable $X_C$ that indicates whether a randomly picked service belongs to the considered category.
    \item We let each metrics $m$ induce a probability distribution $P_m$ over the set of all services, in such a way that the probability of selecting a service is proportional to the value of that metrics for that service.
    \item To measure the importance of knowing a metrics $m$ with respect to a specific category $C$, we compare the distribution of $X_C$ under two different assumptions: that the services are drawn based on $P_C$ and that they are drawn uniformly at random -- the latter meaning that $\Pr[X_C=1]$ is the overall prevalence of $C$ in the graph. 
    \item As a measure of information gain, we use the Kullback-Leibler divergence. 
\end{itemize}
Since the statistical relevance of the above approach relies on a reasonably sized sample, we will only consider the DSGI and DSGU.

\begin{figure*}[htbp]
    \centering
    \begin{subfigure}[b]{0.43\textwidth}
        \includegraphics[width=\textwidth]{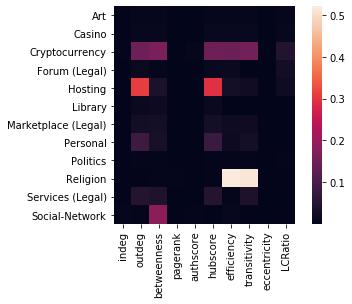}
        \caption{\dcore ``Normal''.}
	    \label{fig:dcore_normal}
    \end{subfigure}
    \begin{subfigure}[b]{0.43\textwidth}
        \includegraphics[width=\textwidth]{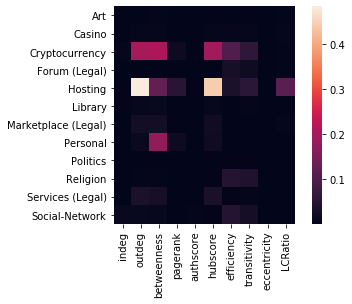}
        \caption{\dunion ``Normal''.}
	    \label{fig:dunion_normal}
    \end{subfigure}
    \begin{subfigure}[b]{0.45\textwidth}
        \includegraphics[width=\textwidth]{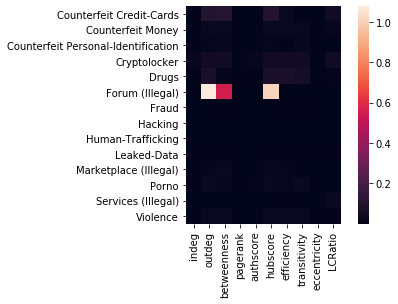}
        \caption{\dcore ``Suspicious''.}
	    \label{fig:dcore_suspicious}
    \end{subfigure}
    \begin{subfigure}[b]{0.45\textwidth}
        \includegraphics[width=\textwidth]{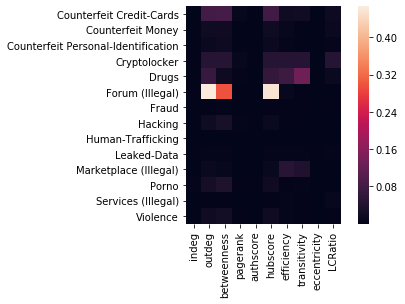}
        \caption{\dunion ``Suspicious''.}
	    \label{fig:dunion_suspicious}
    \end{subfigure}
    
    \begin{subfigure}[b]{0.44\textwidth}
        \includegraphics[width=\textwidth]{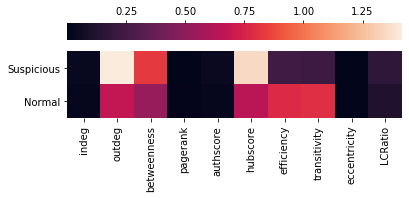}
        \caption{\dcore ``Normal'' \emph{vs.} ``Suspicious''.}
	    \label{fig:dcore_macro}
    \end{subfigure}
    \begin{subfigure}[b]{0.44\textwidth}
        \includegraphics[width=\textwidth]{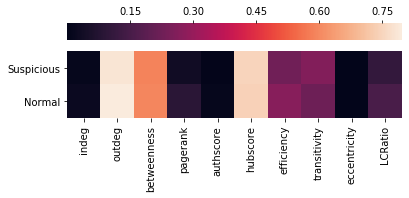}
        \caption{\dunion ``Normal'' \emph{vs.} ``Suspicious''.}
	    \label{fig:dunion_macro}
    \end{subfigure}
    \caption{The information gain provided by different metrics with respect to DUTA classes and macro categories.}
    \label{fig:gain}
\end{figure*}

In Figure~\ref{fig:gain} we show the obtained results, separately considering ``Normal'' classes, ``Suspicious'' classes and their aggregate.
Interestingly, the DSGI and the DSGU broadly provide the same view.
Generally speaking, most of the metrics appear to be uninformative with respect to content-based categories, \emph{i.e.}, the probability of finding a service of a specific class does not increase or decrease significantly when we select the service with probability proportional to most of its topological properties.
However, there are a few remarkable exceptions: (i) the out-degree and the hubscore are especially informative about hosting services and illegal forums; (ii) services discussing religion topics are highlighted by their efficiency and transitivity, arguably because they tend to strongly cluster together; (iii) in the DSGU, the transitivity is also somewhat informative of services that focus on drugs, while the LCRatio is associated with hosting services, even though not as much as one could expect.
These class-level information gains are only partially able to explain the notable improvement that many metrics instead seem to provide to the goal of telling apart, more in general, ``Suspicious'' and ``Normal'' services.
This opens new perspectives towards the design of classifiers that make use of topological features instead of text analysis. 

\section{Conclusions}
\label{sec:conclusion}

We presented an in depth investigation of the key features of the \tor \web graph showing what makes it different from the surface \web graph, inferring on the latent patterns of interactions among \tor users and assessing whether graph metrics can be used to tell apart hidden services playing special roles in the network.
We managed to assess the relationship between contents and structural features and we addressed several open questions about the persistence of the \tor hidden services. 
We discussed the actual changes that may take place in the number and organization of available hidden services and in their inter-connections over time.
In particular we showed that, albeit the \tor \web graph presents a few common features with other real world networks, it has a significantly different structure with respect to the surface \web graph. \tor is a \emph{small world} network composed by a large percentage of volatile hidden services and it is characterized by the presence of in and out-hubs nodes that are critical for the graph connectivity. Network navigation seems to be facilitated only in one direction: users select a starting out-hub and then they move looking for the website of interest. Peripheral nodes, once reached, usually don't provide any possibility to go back and navigate in other directions.
Although many \tor nodes are not persistent, the graph, on the whole, seems to possess a meaningful and stable community structure. 
Finally we showed that, considering a class-level categorization, most of the applied topological metrics appear to be uninformative with respect to the hidden services' content. Nevertheless, some metrics seem to provide a notable improvement in the goal of telling apart ``Suspicious'' from ``Normal'' services.

Future efforts will be devoted to further extending the analysis over time and possibly measuring the influence of exogenous factors ({\em e.g.} changes in the legislation or breaking news from the real-world) on the \tor \web organization.

\bibliographystyle{unsrt}
\bibliography{biblio}

\end{document}